\documentclass[aps,prb,reprint,longbibliography,nofootinbib]{revtex4-2}

\usepackage{amsmath,amssymb,bm,mathtools}
\usepackage{graphicx}
\usepackage[dvipsnames]{xcolor}
\usepackage{microtype}
\usepackage{enumitem}
\usepackage{hyperref}
\usepackage{physics}
\usepackage[normalem]{ulem}

\hypersetup{colorlinks=true,citecolor=MidnightBlue,linkcolor=MidnightBlue,urlcolor=MidnightBlue}
\setlist[itemize]{leftmargin=*,topsep=2pt,itemsep=1pt}

\newcommand{\BZ}{\mathrm{BZ}}

\newcommand{\calAvec}{\bm{\mathcal A}}
\newcommand{\bG}{{\bar g}}
\newcommand{\intk}{\int_{\BZ}\!\frac{ d^2 k}{(2\pi)^2}}
\newcommand{\intkp}{\int_{\BZ}\!\frac{ d^2 k'}{(2\pi)^2}}

\begin{document}

\title{Anomalous Hall Conductivity as an Intrinsic Property of a Tight-Binding Graph }

\author{E. S. Andriyakhina}
\author{G. Lemut}
\author{P. W. Brouwer}
\affiliation{Dahlem Center for Complex Quantum Systems, Fachbereich Physik, and Halle-Berlin-Regensburg Cluster of Excellence CCE, Freie Universit\"at Berlin, Arnimallee 14, 14195 Berlin}

\date{\today}

\begin{abstract}
In a ballistic system, the anomalous dc Hall response generically depends on microscopic details, including the geometric embedding of the orbitals and the precise implementation of the measurement. Here, we consider
the question under what circumstances the anomalous Hall conductivity can be regarded as an intrinsic property of a system. We show that the full dc Hall response $\sigma_{xy}^{\rm H}$ becomes embedding-independent in the presence of relaxation by impurity scattering, although the individual (intrinsic) band and (extrinsic) side-jump and anomalous-distribution contributions to the Hall conductivity separately remain embedding-dependent. We further derive a symmetry criterion for when the band term itself becomes embedding-independent, but argue that this alone neither identifies it with the full Hall conductivity $\sigma_{xy}^{\rm H}$, nor makes it a reliable approximation for $\sigma_{xy}^{\rm H}$. More broadly, our results establish embedding dependence as a diagnostic of whether a response coefficient represents a stationary intrinsic property of a material or retains sensitivity to measurement protocol.
\end{abstract}

\maketitle

\section{Introduction}

The anomalous Hall effect (AHE) is one of the oldest manifestations of spin-orbit-coupled transport in a time-reversal-broken metal. Its modern description combines the band-structure mechanism of Karplus and Luttinger with disorder-induced skew scattering, anomalous-distribution, and side-jump contributions, whose microscopic and semiclassical formulations are now well established~\cite{KarplusLuttinger1954,Smit1955,Berger1970,SinitsynNiuMacDonald2006,Onoda2006,SinitsynEtAl2007,Sinitsyn2008, Kovalev2010, Nagaosa2010}.  The band contribution is commonly written as \cite{Onoda2002}
\begin{equation}
    \sigma_{xy}^{\rm B}=-\frac{e^2}{\hbar}\sum_\lambda\intk f(\varepsilon_{\lambda\bm{k}})\,\Omega_{\lambda,z}(\bm{k}), \label{eq:berryformula}
\end{equation}
where $e>0$ is the elementary charge, $f$ is the Fermi function, and $\Omega_{\lambda,z}$ is the Berry curvature of band $\lambda$.  Equation \eqref{eq:berryformula} is conceptually attractive: it expresses a transport coefficient through the geometry of Bloch wave functions and it connects metallic AHE theory to the quantized Hall response of a band insulator~\cite{TKNN1982,Haldane2004,XiaoChangNiu2010,Nagaosa2010}. It has therefore become a standard bridge between the AHE and the rapidly expanding field of quantum geometry~\cite{YuEtAl2025}.

\begin{figure}[t!]
    \includegraphics[width=0.95\columnwidth]{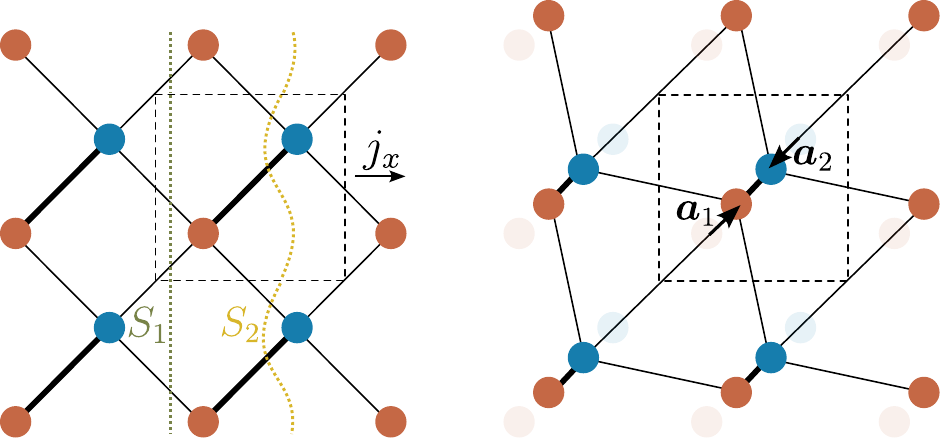}
    \caption{The left and right panels show two geometric embeddings of the same periodic graph, related by a shift of the orbital positions. The spectrum and connectivity of a tight-binding Hamiltonian defined on the two lattices are the same, whereas the microscopic current operator and coupling to the electric field depend on the assigned orbital positions. The dotted square indicates the unit cell. In the left panel, the colored dashed curves ($S_1$ and $S_2$) illustrate different microscopic choices of the current-counting surface for calculating the current density $j_x$ in the horizontal ($x$) direction.
    }
 \label{fig:graph_embedding}
\end{figure}

There is, however, a basic ambiguity hidden in this formula. This ambiguity is most clearly exposed in the context of a tight-binding description of the metal. A tight-binding Hamiltonian specifies hopping amplitudes and on-site terms on an abstract periodic graph. The real-space positions assigned to the orbitals, or the \emph{orbital embedding}, are additional data. Figure \ref{fig:graph_embedding} illustrates an example of two different embeddings of
the same periodic graph with a two-orbital unit cell. Moving orbitals inside the unit cell without changing any hopping amplitude leaves the spectrum, band topology, and graph connectivity unchanged, but changes the position operator, the current operator, the microscopic coupling to an electric field, and the local-in-$\bm{k}$ Berry curvature \cite{SimonRudner2020}. This distinction between geometry-dependent and geometry-independent quantities was emphasized in an article by Simon and Rudner~\cite{SimonRudner2020}; embedding-related issues also arise in polarization, topology, Wannier representations, superfluid response, and the definition of quantum geometry itself~\cite{MarzariEtAl2012,FuchsPiechon2021,TamPeotta2024,OhMurakami2026,WangRacaSimon2026,TelleLange2026}.

For two different embeddings of the same graph (as shown, {\em e.g.}, in Fig.\ \ref{fig:graph_embedding}), let the positions of the orbital $q$ differ by $\bm{a}_q$, where $q = 1, 2, \dots, N_{\rm orb}$ labels an orbital, spin, sublattice, or other internal degree of freedom. In momentum space, the two embeddings are related by a unitary transformation of the Bloch Hamiltonian $h_0(\bm{k})$ generated by the diagonal matrix $U_{\bm k}=\text{diag}\left(e^{-i\bm{k}\cdot\bm{a}_1},\dots,e^{-i\bm{k}\cdot\bm{a}_{N_{\rm orb}}}\right)$. Simon and Rudner then pointed out that the Berry curvature of band $\lambda$ in the two embeddings differs by \cite{SimonRudner2020}
\begin{equation}
  \Delta \bm{\Omega}_{\lambda}(\bm{k}) = \nabla_{\bm{k}} \times \sum_q \bm{a}_q w_{\lambda q}(\bm{k}),
  \label{eq:Delta_berryformula}
\end{equation}
where $w_{\lambda q}(\bm{k})=|\braket{q}{u_{\lambda\bm{k}}}|^2$ is the weight of orbital $q$ in the Bloch state $|u_{\lambda\bm{k}}\rangle$.
For a completely filled isolated band, the embedding dependence of the local Berry curvature in Eq.~\eqref{eq:Delta_berryformula} integrates to zero, leaving the Chern number and the zero-temperature Hall conductance invariant. For a metal, only part of a band contributes to the integration in Eq.~\eqref{eq:berryformula} and no such cancellation occurs: In general, the band contribution $\sigma^{\rm B}_{xy}$ to the AHE of Eq.~\eqref{eq:berryformula} is embedding-dependent. 

On the other hand, in the presence of disorder, it is well known that the Berry-curvature term $\sigma^{\rm B}_{xy}$ is only one contribution to the AHE, alongside side-jump, anomalous-distribution, and skew-scattering terms that can be comparable in magnitude~\cite{SinitsynNiuMacDonald2006,Onoda2006,SinitsynEtAl2007,Sinitsyn2008,Nagaosa2010,Kovalev2010}. In view of the embedding dependence of $\sigma^{\rm B}_{xy}$ it is relevant to ask about the embedding dependence of the other contributions to the Hall conductivity and about a possible embedding dependence of the full Hall response $\sigma^{\rm H}_{xy}$. If $\sigma_{xy}^{\rm H}$ is embedding-independent, the Hall conductivity is a property of the abstract periodic graph Hamiltonian, without necessity to specify the positions of the orbitals inside the unit cell. We will find that the answer to the question of embedding dependence is closely related to the question whether the transverse current response to an applied electric field depends on details that are typically beyond experimental control, such as the microscopic cross section through which current is measured or the spatial profile of the applied electric field within a unit cell. 

In this article, we show that
\begin{enumerate}[label=(\roman*)]
    \item The Hall response is generically embedding-dependent in the ballistic regime (relaxation time for impurity scattering $\tau \to \infty$, before taking the limit that the driving frequency $\omega \to 0$). This dependency signals that, without relaxation, a uniform static electric field does not define a unique stationary bulk transport problem and that the Hall response cannot meaningfully be characterized by a ``Hall conductivity'' $\sigma^{\rm H}_{xy}$.

    \item In the presence of disorder, the Hall conductivity $\sigma_{xy}^{\rm H}$ is embedding-independent in the relaxation-dominated dc limit (frequency $\omega \to 0$ at fixed, finite relaxation time $\tau$). Of the four terms of the decomposition of $\sigma_{xy}^{\rm H}$ into an ``intrinsic'' band contribution and ``extrinsic'' side-jump, anomalous-distribution, and skew-scattering terms, only the skew-scattering term is separately invariant under a change of embedding. For a model with uncorrelated on-site disorder, we verify by explicit calculation that embedding independence is already exact within the non-crossing approximation.
    
    \item
    Although the band contribution $\sigma^{\rm B}_{xy}$ of 
    Eq.\ (\ref{eq:berryformula}) is embedding-dependent for a generic tight-binding Hamiltonian, symmetry constraints can enforce that the Berry-curvature integral of Eq.\ (\ref{eq:berryformula}) is embedding-independent for specific cases. 

    \item One may seek an ``optimal'' embedding for which the band contribution reproduces the full Hall conductivity, $\sigma^{\rm B}_{xy}=\sigma_{xy}^{\rm H}$. This is possible in principle if the band contribution is itself embedding-dependent. However, the ``optimal'' embedding for which $\sigma^{\rm B}_{xy} = \sigma_{xy}^{\rm H}$  generally varies with the filling fraction and with the orbital structure of the disorder, so it does not define a unique property of the graph or the tight-binding Hamiltonian defined on it.
\end{enumerate}

The issue of embedding dependence or independence is distinct from the freedom to choose a periodic or atomic Bloch basis~\cite{Bena2009,Cayssol2021,Esteve-Paredes2023}. For fixed positions of the lattice sites, these are merely different representations of the same physical problem, and consistently transformed operators give identical responses. The corresponding intracell velocity correction \cite{Esteve-Paredes2023,Boykin2001} restores representation covariance, but it neither removes the dependence on the physical position operator nor establishes stationarity. Conversely, once embedding independence of the complete dc response in the presence of relaxation has been established, the calculation may be performed in any convenient embedding, including one with coincident intracell orbitals, the so-called ``periodic gauge''.\footnote{The embedding for which all orbitals in the unit cell are assigned the same position is called this way, because the Bloch Hamiltonian $h_0(\bm k)$ is a periodic function of $\bm k$ for this choice of embedding.}

Our findings are corroborated in the remainder of this article. In Sec.\ \ref{sec:graph} we analyze how the conductivity tensor $\sigma_{\beta\alpha}$ of a generic periodic tight-binding Hamiltonian changes under a change of embedding. We relate the embedding dependence to a dependence of the conductivity on the choice of the cross section for measuring the current and the precise spatial profile of the applied electric field within the unit cell and we present general arguments for the conclusions (i) and (ii) listed above. Section~\ref{sec:berry} analyzes Eq.~\eqref{eq:berryformula}, discusses the conditions for embedding independence of the band contribution $\sigma_{xy}^{\rm B}$ and the role of spatial symmetry, and addresses the possibility of an embedding that is optimized for the calculation of the anomalous Hall conductivity. We consider a tight-binding Hamiltonian with uncorrelated diagonal disorder in Sec.\ \ref{sec:disorder} and show how relaxation by impurity scattering and the non-crossing Ward identity ensure that the Hall conductivity $\sigma_{xy}^{\rm H}$ is embedding-independent in a calculation that treats the disorder in the non-crossing approximation. In Sec.~\ref{sec:numerics} we illustrate our findings for a simple two-band model. We conclude the article with a discussion of our results in Sec.\ \ref{sec:discussion}. Derivations and further numerical results are collected in the Appendices.

\section{Graph, embedding, and coupling to the electric field}
\label{sec:graph}

We consider a periodic graph with nodes $\bm{s}=(\bm n,q)$, where $\bm n$ labels a Bravais cell and $q=1,\ldots,N_{\rm orb}$ labels the orbitals inside a unit cell. A single-particle tight-binding Hamiltonian may be written as
\begin{equation}
    H= H_0 + V,
 \label{eq:graphH}
\end{equation}
with
\begin{align}
  H_0 =&\, -\sum_{\bm{s}\neq \bm{s}'}c_{\bm{s}}^\dagger t_{\bm{s}\bm{s}'}c_{\bm{s}'}+\sum_{\bm{s}} c_{\bm{s}}^\dagger u_{0\bm{s}} c_{\bm{s}}, \\
  V =&\, \sum_{\bm{s}} c_{\bm{s}}^\dagger u_{\bm{s}} c_{\bm{s}},
\end{align}
where $t_{{\bm{s}}{\bm{s}}'} = t_{{\bm{s}}'{\bm{s}}}^*$ are hopping amplitudes and $u_{0{\bm{s}}}$ and $u_{{\bm{s}}}$ are onsite potentials. The hopping amplitudes and the on-site potential $u_{0{\bm{s}}}$ are periodic functions of the node index ${\bm{s}}$; the potential $u_{{\bm{s}}}$ is a random potential, which breaks the lattice periodicity. 

Formally, the Hamiltonian of Eq.~\eqref{eq:graphH} is defined on a purely combinatorial object: a set of nodes and edges, periodic under discrete translations, with no notion of distance, angle, or position other than the periodicity. This set becomes a lattice only after assigning positions
\begin{equation}
    \bm{r}_{\bm n q}=\sum_{j = 1}^2 n_j \hat{e}_j +\bm{\rho}_q
\label{eq:embedding}
\end{equation}
to the nodes of the graph, where $\hat{e}_j$, $j=1,2$ is a basis vector of the $2$-dimensional Bravais lattice and $\bm{\rho}_q$ a vector that indicates the position of the node $q$ relative to the center of a unit cell. 
The assignment of the positions in Eq.\ (\ref{eq:embedding}) is what is referred to as the {\em geometric embedding} of the graph. Figure \ref{fig:graph_embedding} shows an example of two different embeddings for the same periodic graph with a two-orbital unit cell.

We will consider embeddings $\bm{r}_{{\bm n}q}$ for the same graph that differ by the choice of the vectors ${\bm{\rho}}_q$, while keeping the basis vectors $\hat{e}_j$ of the Bravais lattice the same. The reason for keeping the basis vectors the same is that these enter in macroscopic properties, such as the relation between density and filling fraction or between electric field and bias voltage. 

The conductivity $\sigma_{\beta\alpha}$ describes the linear response of the current density $\bm{j}$ to an applied electric field $\bm{E}$,
\begin{equation}
j_{\beta} = \sum_{\alpha} \sigma_{\beta\alpha} E_{\alpha}.
\label{eq:jlin}
\end{equation}
The current density $j_{\beta}$ is defined in terms of the charge flowing through a periodic reference cross section; see the left panel of Fig.~\ref{fig:graph_embedding} for two examples of cross sections that can be used to measure the current density in the $x$ direction of the graph. The precise position and shape of this reference cross section relative to the center of a unit cell can be chosen freely. Rather than using a single reference cross section, $\bm{j}$ is often defined in the literature by averaging the currents through several such cross sections. Similarly, implementing the electric field $\bm{E} = \sum_{j} E_j \hat e_j$ requires a periodic array of surfaces across which the potential drops by $\bm{E}\cdot\hat e_j$. Again, there is freedom in the placement of these surfaces within the unit cell, and the potential drop is often distributed over several steps.

The conductivity tensor $\sigma_{\beta\alpha}$ is an intrinsic property of the graph if the current response is independent of the choice of reference surfaces used to define $\bm{j}$ and $\bm{E}$. By contrast, a dc response that depends on these choices is of limited experimental relevance, since it depends on microscopic details that are not under experimental control. A realistic dc transport experiment requires attaching current-carrying leads and voltage probes, and a ``conductivity'' can be inferred from such an experiment only if it is insensitive to the precise details of how and where these contacts are attached to the sample.

We will now show that independence with respect to choice of reference surfaces for $\bm{j}$ and $\bm{E}$ is equivalent to being independent with respect to the choice of the geometric embedding.
To see explicitly where the embedding enters the transport problem and how it is related to the choice of the reference surfaces, we consider a gauge in which a spatially uniform electric field
\begin{equation}
    \bm{E}(t)=\bm{E}_0\cos(\omega t)
\end{equation}
is represented by a spatially uniform vector potential
\begin{equation}
    \bm{A}(t)=-\frac{c\bm{E}_0}{\omega}\sin(\omega t).
\end{equation}
We take the limit $\omega \to 0$ at the end of the calculation.
The vector potential couples to the tight-binding model of Eq.~\eqref{eq:graphH} through the Peierls substitution: A hopping amplitude $t_{ss'}$ acquires the phase
\begin{equation}
    t_{{\bm{s}}{\bm{s}}'}\rightarrow t_{{\bm{s}}{\bm{s}}'}(\bm{A}) = t_{{\bm{s}}{\bm{s}}'} 
    e^{\frac{ie}{\hbar c} \bm{A}(t)\cdot \left(\bm{r}_{\bm{s}}-\bm{r}_{{\bm{s}}'}\right)}.
\label{eq:t_Peierls}
\end{equation}
Thus, whereas the zero-field tight-binding Hamiltonian can be defined purely in terms of the graph, its coupling to an electric field depends explicitly on the positions assigned to the orbitals.

The same geometric information also enters the definition of the current density.  Denoting the Hamiltonian after the Peierls substitution by $H(\bm{A})$, one may define the  current-density operator by differentiating with respect to the uniform vector potential \cite{Graf1995},
\begin{equation}
    \bm{j}(\bm{A}) = -\frac{c}{L^2} \frac{\partial H(\bm{A})}{\partial \bm{A}}, \label{eq:jdef}
\end{equation}
where $L^2$ is the volume of the system. Since the Peierls phase is applied on all bonds, taking the derivative in Eq.~\eqref{eq:jdef} yields a sum of bond currents weighted by their spatial displacements. For the $x$ component, this weighting corresponds to averaging the current through vertical reference cross sections whose positions are sampled uniformly over one unit cell, connecting Eq.~\eqref{eq:jdef} to the cross-section picture introduced above. Since $H(\bm{A})$ depends on the displacement vectors $\bm{r}_{\bm{s}}-\bm{r}_{{\bm{s}}'}$, the current operator $\bm{j}(\bm{A})$ also depends on the chosen embedding. This observation is central to what follows.

Two embeddings may describe exactly the same graph and therefore have the same zero-field spectrum, while nevertheless producing different microscopic couplings to the electric field and different current operators. In the special case in which all nodes within a unit cell are assigned the same position --- referred to as the \emph{periodic gauge}, since $H(\bm{A})$ is then a periodic function of the vector potential $\bm{A}$~\cite{Zak1989,Resta2000,Cayssol2021} --- the Hamiltonian $H(\bm{A})$ corresponds to an implementation of the electric field in which the potential drops only between unit cells and not within a unit cell. Correspondingly, the current operator $\bm{j}(\bm{A})$ gives the current flowing through the a reference surface at the boundary between unit cells. A generic embedding, in which each node $\bm{s}$ has its own spatial position $\bm{r}_{\bm{s}}$, then corresponds to a current density $\bm{j}$ that is the average of the currents through multiple such cross sections.

We compare two embeddings related by a shift of the orbital positions within the unit cell,
\begin{equation}
    \widetilde{\bm{\rho}}_{q} = \bm{\rho}_{q}+\bm{a}_q. \label{eq:shift}
\end{equation}
The underlying tight-binding graph is unchanged by this transformation; only the real-space positions assigned to its orbitals are modified. Denoting the Hamiltonians obtained after the Peierls substitution~\eqref{eq:t_Peierls} in the two embeddings by $H(\bm A)$ and $\widetilde H(\bm A)$, respectively, one finds \cite{Cayssol2021}
\begin{equation}
    \widetilde H(\bm A) = e^{i e \bm A\cdot\bm a/\hbar c} H(\bm A) e^{-i e \bm A\cdot\bm a/\hbar c},
\end{equation}
where $\bm a$ is diagonal in the orbital basis and assigns the displacement $\bm a_q$ to orbital $\bm{n}q$. In first-quantized notation, its matrix elements are $(\bm a)_{\bm n q,\bm n' q'}=\bm a_q\delta_{\bm n\bm n'}\delta_{qq'}$.

It is useful to introduce the density associated with orbital $q$,
\begin{equation}
    n_q = \frac{1}{L^2}\sum_{\bm n}c_{\bm n q}^{\dagger}c_{\bm n q},
\label{eq:nq}
\end{equation}
so that the change in polarization between the two embeddings is simply $\sum_q \bm a_q n_q$. At time $t=0$, the vector potential $\bm{A}(0) = 0$, so that $H=\widetilde H$ and the current operators defined by Eq.~\eqref{eq:jdef} are related by \cite{Boykin2001}
\begin{align}
    \widetilde{\bm j} - \bm j =&\, -\frac{i e}{\hbar} \sum_q \bm{a}_q [n_q,H] \notag \\ =&\, e \sum_q \bm{a}_q \frac{dn_q}{dt},
 \label{eq:jtransform}
\end{align}
where in the second line we used the Heisenberg equation of motion for $n_q$.
Thus, changing the embedding changes the current operator by the time derivative of the polarization associated with the orbital displacements.

The embedding also affects the way the external electric field enters the Hamiltonian. To see this, one can undo the embedding-dependent part of the vector-potential coupling by a time-dependent unitary transformation. Using $\bm E=-(1/c)\partial_t\bm A$, the Hamiltonian in the shifted embedding is mapped to
\begin{equation}
    \widetilde H(\bm A)\to H(\bm A)-eL^2\sum_q \bm E\cdot\bm a_q\,n_q. \label{eq:fieldtransform}
\end{equation}
Hence, a change of embedding appears as an additional spatially periodic scalar potential. A change of embedding therefore affects the transport problem in two distinct but related ways: it modifies both the microscopic definition of the current and the microscopic realization of the applied electric field.

We can now state the physical condition for the dc conductivity to be independent of the embedding.\footnote{The ac conductivity at finite frequency $\omega$ is always embedding-dependent.}  According to Eqs.~\eqref{eq:jtransform} and~\eqref{eq:fieldtransform}, this requires two closely related properties:
\begin{enumerate}[label=(\roman*)]
    \item the change in polarization $\sum_q \bm a_q n_q$ induced by a static uniform electric field must itself become stationary, so that its time derivative $\dot n_q$ does not contribute to the dc current;
    \item the additional periodic potential $-eL^2 \sum_q \bm E\cdot\bm a_q\,n_q$ associated with a change of embedding must not by itself drive a stationary current.
\end{enumerate}

In a system with a finite relaxation time, both conditions are naturally satisfied: after the external electric field is switched on, the electronic system relaxes toward a stationary state and the polarization response saturates at long times, whereas a periodic change of the potential causes a transient current response, but no steady-state current flow.

The situation is qualitatively different in a ballistic system.  In the absence of relaxation, the current and polarization response to a static electric field do not approach a stationary limit, but typically grow linearly with time. The time derivative of the polarization then remains finite in the long-time limit, and the currents obtained from two different embeddings need not agree. The same conclusion applies to the response to a periodic potential: whereas with relaxation a periodic potential only gives a transient current response, the current response typically does not decay if there is no relaxation, leaving a possibly nonzero steady-state current response. 

To formalize these conclusions, we introduce the orbital-resolved susceptibility $\chi_{q\alpha}(\omega)$, which describes the linear response of the density $n_q$ to an applied electric field
\begin{equation}
    \delta n_q(\omega)=\sum_{\alpha}\chi_{q\alpha}(\omega)E_{\alpha}(\omega).
 \label{eq:chi}
\end{equation}
Within the Drude theory, the same susceptibility $\chi_{q\alpha}(\omega)$ also governs the current response to a periodic potential. From this observation, Eq.~\eqref{eq:jtransform}, and the equation of motion for $n_q$, it follows that the difference between Hall conductivities for the two embeddings is
\begin{equation}
    \Delta\sigma_{\beta\alpha}(\omega) = -i e\omega\sum_q \left[(a_q)_{\beta}\chi_{q\alpha}(\omega) -(a_q)_{\alpha}\chi_{q\beta}(\omega)\right].
\label{eq:masteridentity}
\end{equation}
The two terms in brackets arise, respectively, from the change \eqref{eq:jtransform} of the current operator and from the change \eqref{eq:fieldtransform} of the coupling to the electric-field. 

The role of relaxation can be clearly identified from Eq.~\eqref{eq:masteridentity}. Without disorder, the leading (Drude) intraband response is
\begin{equation}
    \chi_{q\alpha}^{(0)}(\omega) = 
    e \sum_{\lambda} \int \frac{d^2 k}{(2 \pi)^2}
    \left(-\frac{d f(\varepsilon_{\lambda \bm{k}})}{d\varepsilon_{\lambda \bm{k}}} \right) 
    \frac{w_{\lambda q}(\bm{k})v_{\lambda \alpha}(\bm{k})}{-i\omega+\eta},
 \label{eq:chiballistic}
\end{equation}
where 
\begin{equation}
    w_{\lambda q}(\bm{k}) = |\braket{q}{u_{\lambda \bm{k}}}|^2
\end{equation}
is the orbital weight of the Bloch state $\ket{u_{\lambda \bm{k}}}$ of the $\lambda $-th band and 
\begin{equation}
  \label{eq:group_velocity}
  \bm v_\lambda(\bm{k}) = \frac{1}{\hbar} \nabla_{\bm{k}}\varepsilon_{\lambda \bm{k}}
\end{equation}
the corresponding group velocity. The factor $\omega$ in Eq.~\eqref{eq:masteridentity} is canceled by the denominator in Eq.\ (\ref{eq:chiballistic}), leaving a finite embedding dependence in the dc limit $\omega \to 0$. If momentum and orbital polarization relax on a finite time scale $\tau$, the ballistic pole is cut off for $\omega \sim 1/\tau$. Then $\chi_{q\alpha}(0)$ is finite for $\omega \to 0$ and $\Delta \sigma_{\beta\alpha}(\omega) \to 0$ as $\omega\to 0$.  

The physical conclusion is simple:  A ``conductivity tensor'' $\sigma_{\beta\alpha}$ is meaningful only if it is embedding-independent. In general, an embedding-independent dc conductivity requires a stationary response to an applied electric field.

\section{The Berry-curvature formula in a ballistic metal}
\label{sec:berry}

\subsection{Conductivity from the Kubo formula}

We first analyze the transverse current response in the absence of the disorder potential $V$. 
For a fixed embedding, the lattice translation invariance allows the Hamiltonian $H_0$ to be block-diagonalized in reciprocal space,
\begin{equation}
    H_0 = \sum_{\bm{k}} \sum_{q,q'} c^{\dagger}_{{\bm k},q} h_{0qq'}(\bm{k}) c_{{\bm k},q'},    \label{eq:hvelocity}
\end{equation}
with $c_{\bm{k},q} = L^{-1} \sum_{\bm{n}} c_{\bm{n}q} e^{-i \bm{k} \cdot \bm{r}_{\bm{n}q}}$. (Recall that the lattice positions $\bm{s}$ are written $\bm{s} = (\bm{n},q)$, see Sec.\ \ref{sec:graph}.)
The Bloch Hamiltonian $h_0({\bm{k}})$ is an $N_{\rm orb}\times N_{\rm orb}$ matrix, which admits the spectral decomposition
\begin{equation}
    h_0(\bm{k}) = \sum_\lambda \ket{u_{\lambda\bm{k}}} \varepsilon_{\lambda\bm{k}}\bra{u_{\lambda\bm{k}}},
 \label{eq:hvelocity_diag}
\end{equation} 
where the Bloch eigenstate $\ket{u_{\lambda\bm{k}}}$ is an $N_{\rm orb}$-component vector.

For the same embedding we also introduce the electric field $\bm{E}$, define the current density $\bm{j}$, and the conductivity as described in Sec.\ \ref{sec:graph}. The finite-frequency conductivity tensor $\sigma_{\beta\alpha}(\omega)$ can be expressed in terms of single-particle Green functions as \cite{Kubo1957,Bastin1971,CrepieuxBruno2001}
\begin{align}
  \label{eq:kubo}
    \lefteqn{\sigma_{\beta\alpha} (\omega) =\, -e^2 \int \frac{dE}{2\pi\omega}\, f(E) 
    \intk} ~~~~~
    \nonumber \\ &\, \mbox{} \times
    \tr \Bigg[ 
    \frac{1}{\hbar} 
    \frac{\partial v_{\bm{k}\beta}}{\partial k_\alpha} 
    ( g^R(\bm{k},E) -g^A(\bm{k},E) ) 
    \\ &\, \ \ \mbox{}
    + v_{\bm{k}\beta} g^R(\bm{k},E+\hbar\omega) 
    v_{\bm{k}\alpha} 
    ( g^R(\bm{k},E) - g^A(\bm{k},E)) 
    \nonumber \\ \nonumber &\, \ \ \mbox{}
    + v_{\bm{k}\beta} 
    ( g^R(\bm{k},E) - g^A(\bm{k},E))
    v_{\bm{k}\alpha} 
    g^A(\bm{k},E-\hbar\omega) 
    \Bigg].
\end{align}
The trace is taken over the orbital degrees of freedom and the momentum integral extends over the Brillouin zone. The velocity operator
\begin{equation}
    \bm{v}_{\bm{k}} = \frac{1}{\hbar} \bm{\nabla}_{\bm{k}} h_0(\bm{k})
\label{eq:bloch_velocity}
\end{equation}
and retarded and advanced Green functions
\begin{equation}
    g^{R/A}(\bm{k},E) = \left( E-h_0(\bm{k}) \pm i\eta \right)^{-1},
\label{eq:clean_green_functions}
\end{equation}
are $N_{\rm orb}\times N_{\rm orb}$ matrices. 

The velocity operator $\bm{v}$ contains both intraband and interband contributions. The intraband contribution reads
\begin{equation}
  \label{eq:vintra}
  \bm{v}^{\rm intra} = \sum_{\lambda} \ket{u_{\lambda \bm{k}}} \bm{v}_{\lambda\bm{k}} \bra{u_{\lambda \bm{k}}},
\end{equation}
where the diagonal velocity matrix element $\bm{v}_{\lambda\bm{k}} = \bra{u_{\lambda\bm{k}}} \bm{v}_{\bm{k}} \ket{u_{\lambda\bm{k}}}$
is the group velocity of band $\lambda$, see Eq.\ (\ref{eq:group_velocity}). The interband contribution is
\begin{equation}
  \label{eq:vinter}
  \bm{v}^{\rm inter} = \bm{v} - \bm{v}^{\rm intra}.
\end{equation} 
The leading singular contribution in the low-frequency limit arises from the intraband response and yields the symmetric Drude conductivity
\begin{align}
    \sigma_{\beta\alpha}^{\rm sym}(\omega)
    &=e^2 \sum_{\lambda}\intk
    \left( -\frac{df(\varepsilon_{\lambda \bm{k}})}{d \varepsilon_{\lambda \bm{k}}}\right) \frac{(v_{\lambda \bm{k}})_\alpha(v_{\lambda \bm{k}})_\beta}{-i\omega}.
\label{eq:drude_conductivity}
\end{align}
The antisymmetric interband Kubo response from Eq.~\eqref{eq:kubo} reduces to Eq.~\eqref{eq:berryformula}~\cite{Kubo1957,Bastin1971,CrepieuxBruno2001,Nagaosa2010}. 
Unlike the Drude conductivity, which diverges $\propto \omega^{-1}$ for $\omega \to 0$, the Hall conductivity obtained from Eq.\ (\ref{eq:kubo}) remains finite for $\omega \to 0$.

\subsection{Embedding dependence of the Hall response}

We now consider the change of the conductivity $\sigma_{\beta\alpha}$ under a change of embedding $\bm{\rho}_{q} \to \tilde{\bm{\rho}}_{q} = \bm{\rho}_{q} + \bm{a}_q$, see Eq.\ (\ref{eq:shift}). We arrange the shift vectors $\bm{a}_q$ into a $N_{\rm orb} \times N_{\rm orb}$ matrix
\begin{equation}
  \label{eq:adef}
  \bm{a} = \sum_{q} \bm{a}_q P_q,
\end{equation}
where $P_q = \ket{q} \bra{q}$ is the projector onto the orbital $q$. With this notation, under a change of embedding the Bloch Hamiltonian $h_0(\bm{k})$ changes as \cite{Dobardzic2014,SimonRudner2020}
\begin{equation}
  \label{eq:hchange}
  h_0(\bm{k}) \to \tilde h_0(\bm{k}) =
  U_{\bm{k}}
  h_0(\bm{k}) 
  U_{\bm{k}}^{\dagger},
\end{equation}  
where
\begin{equation}
  \label{eq:Udef}
    U_{\bm{k}} = 
    e^{-i {\bm k} \cdot {\bm a}}.
\end{equation}
From Eq.~\eqref{eq:hchange}, we conclude that the band energies $\varepsilon_{\lambda\bm{k}}$ are left unchanged, whereas the periodic Bloch eigenstates transform as 
\begin{equation}
  \label{eq:tildeuchange}
  \ket{u_{\lambda\bm{k}}}\to \ket{\widetilde{u_{\lambda\bm{k}}}} =  U_{\bm{k}}
  \ket{u_{\lambda\bm{k}}}.
\end{equation}  
Since both the group velocities and the Drude conductivity depend only on the band energies, a change of embedding leaves them unchanged. By contrast, because it depends explicitly on the Bloch eigenstates, the antisymmetric interband response generally changes under a change of embedding. 

From Eq.~\eqref{eq:tildeuchange}, we find that under a change of embedding the Berry connection ${\calAvec}_\lambda(\bm{k}) =\bra{{u_{\lambda\bm{k}}}} i \nabla_{\bm{k}} \ket{{u_{\lambda\bm{k}}}}$
changes according to
\begin{equation}
    \widetilde{\calAvec}_\lambda (\bm{k}) - \calAvec_\lambda (\bm{k}) = \sum_q\bm{a}_q w_{\lambda q}(\bm{k}).\label{eq:Atransform}
\end{equation}
Equation~\eqref{eq:Atransform} gives the change $\Delta \bm{\Omega}_{\lambda}(\bm{k})$ of the Berry curvature in Eq.~\eqref{eq:Delta_berryformula}. The Chern number, which contains the integral of $\Omega_{\lambda,z}(\bm{k})$ over a full band, is invariant under a change of the embedding, because the Brillouin-zone integral of $\Delta \Omega_{\lambda,z}(\bm{k})$ vanishes.  A partially occupied band is different.  Integrating Eq.~\eqref{eq:berryformula} with Eq.~\eqref{eq:Delta_berryformula} by parts yields
\begin{align}
 \Delta\sigma_{xy}^{\rm B} = - e^2 \sum_q\bm{a}_q\cdot\bm{\Xi}_q(\mu,T),
 \label{eq:DeltasigmaXi}
\end{align}
where 
\begin{align}
    \bm{\Xi}_q(\mu,T) = \hat{\bm z}\times 
    \sum_\lambda \intk \left(-\frac{df(\varepsilon_{\lambda\bm{k}})} {d\varepsilon_{\lambda\bm{k}}}\right) 
  {\bm{v}}_{\lambda \bm{k}} w_{\lambda q}(\bm{k})
 \label{eq:Xi}
\end{align}
is an embedding-independent vector.
Note that, whereas the band contribution $\sigma^{\rm B}_{xy}$ to the Hall conductivity contains an integration over the entire Brillouin zone, its embedding dependence is entirely controlled by the Fermi surface, consistent with the Fermi-surface formulation of the metallic intrinsic AHE~\cite{Haldane2004}.

\begin{figure}[t]
    \includegraphics[width=0.95\columnwidth]{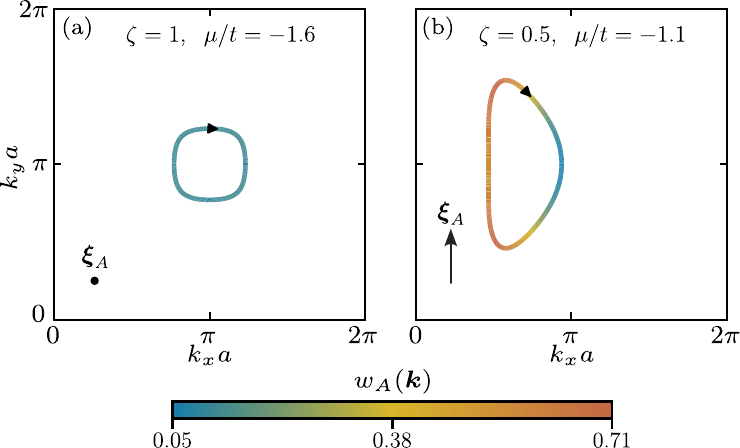}
    \caption{Orbital weights $w_{\lambda A}(\bm{k}) = 1 - w_{\lambda B}(\bm{k})$ at the Fermi energy $\varepsilon_{\lambda \bm{k}} = \mu$ for the two-band Chern-insulator model discussed in Sec.~\ref{subsec:twoband}, 
    where $A$ and $B$ label the two orbitals. The integral $\xi_A(E)$ of $w_{\lambda A}$ around the oriented Fermi contour determines whether the band contribution $\sigma^{\rm B}_{xy}$ to the Hall conductivity is embedding-dependent, see Eq.\ (\ref{eq:xi}). In panel (a), we set $\zeta=1$ and $\mu/t=-1.6$, so that inversion symmetry enforces $\bm{\xi}_A(E) =0$, making $\sigma^{\rm B}_{xy}$ insensitive to the embedding. In the panel (b), this symmetry is broken by setting $\zeta=0.5$, making $\bm{\xi}_A(E)$ nonzero and $\sigma_{xy}^{\rm B}$ embedding-dependent. We choose $\mu/t=-1.1$ for the panel (b). The lattice constant is set to 1 for both panels.}
 \label{fig:xi_concept}
\end{figure}

It is useful to resolve Eq.~\eqref{eq:Xi} in energy,
\begin{align}
    \bm{\Xi}_q(\mu,T) = \int d E
    \left( - \frac{df(E)}{dE}\right)
    \bm{\xi}_q(E),
 \label{eq:Xiconv}
\end{align}
where 
 \begin{align}
    \bm{\xi}_q(E) = \frac{1}{\hbar} \sum_\lambda\oint_{\varepsilon_{\lambda\bm{k}}=E}\frac{w_{\lambda q}(\bm{k})\,d\bm{k}}{(2\pi)^2}
 \label{eq:xi}
\end{align}
is an oriented line integral along the constant-energy contour on the Brillouin-zone torus, see Fig.~\ref{fig:xi_concept} for an example. The orientation of each constant-energy contour is chosen such that $d\bm{k}=(\hat{\bm z}\times\bm v_{\lambda\bm k}/|\bm v_{\lambda\bm k}|)\,d\ell$, where $d\ell>0$ is the arc-length element. Thus, contours surrounding electron and hole pockets have opposite orientations, and the integral includes all components of the constant-energy contour. This vector has an intuitive meaning: $\bm{\xi}_q(E)$ measures how the orbital weight is distributed along the directed Fermi contour.  A constant orbital weight gives zero on any closed contour; embedding sensitivity requires a correlation between orbital texture and the reciprocal-space direction along the contour. 

Completeness implies
\begin{equation}
    \sum_q\bm{\xi}_q(E)=0, \label{eq:xisumzero}
\end{equation}
so only relative orbital displacements matter. Equation~\eqref{eq:xisumzero} is also the statement that translating all orbitals by the same vector cannot change a Berry curvature.

\subsection{Symmetry constraints on embedding dependence \label{sec:class}}

When is the band contribution to the Hall conductivity independent of the choice of the embedding?
From Eqs.~\eqref{eq:DeltasigmaXi}--\eqref{eq:xi} we immediately conclude that the band contribution $\sigma^{\rm B}_{xy}$ is independent of the choice of the embedding for all filling fractions and temperatures if and only if
\begin{equation}
    \bm{\xi}_q(E)\equiv0 \quad \text{$\forall q$ and every energy $E$}.
 \label{eq:classcriterion}
\end{equation}
This is a weaker condition than requiring momentum-independent orbital weights: the weighted contour integral may vanish by cancellation between different sectors of the Fermi contour even when the orbital texture varies strongly along the contour.

A spatial symmetry may provide a sufficient condition, in which such an embedding independence of $\sigma^{\rm B}_{xy}$ will occur. If a point-group operation $g$ maps a constant-energy orbit onto itself, acts in momentum space by $R_g$, and maps the orbital $q \to gq$, then
\begin{equation}
    \bm{\xi}_{gq}(E) = \det(R_g)R_g\bm{\xi}_q(E).
 \label{eq:xisymmetry}
\end{equation}
Hence, if $g$ leaves the orbital sector invariant, so that $gq=q$, we see that rotations and inversion force $\bm{\xi}_q(E)$ to be zero for all energies. Mirror reflection generically removes only one component of $\bm{\xi}_q$, so a single mirror symmetry is in general not sufficient to enforce embedding independence. A symmetry operation that exchanges orbitals is less restrictive. For example, for two orbitals it may only imply $\bm\xi_1=-\bm\xi_2$, without forcing either vector to vanish. The general symmetry statement is derived in Appendix~\ref{app:symmetry}.

In the presence of disorder, there are additional contributions to the Hall conductivity related to impurity scattering. These may be nonzero and of comparable magnitude as the band contribution~\cite{Nagaosa2010, Berger1970, Smit1955, CrepieuxBruno2001, SinitsynEtAl2007, Sinitsyn2008}, even if the condition of Eq.\ \eqref{eq:classcriterion} is satisfied.

\subsection{Why there is no ``optimal'' embedding with $\sigma^{\rm B}_{xy} = \sigma^{\rm H}_{xy}$ in general}
\label{sec:optimal}

In principle, the embedding dependence of the band contribution $\sigma^{\rm B}_{xy}$ to the Hall conductivity can be used to one's advantage, by choosing the embedding in such a way that the band contribution $\sigma^{\rm B}_{xy}$ equals the full Hall conductivity $\sigma_{xy}^{\rm H}$. Writing the displacement vectors of this special ``optimal'' embedding (defined with respect to a reference embedding) as $\bm{a}^{\rm o}$, the condition
\begin{equation}
    \sigma_{xy}^{\rm B}(\mu,T; \bm a^{\rm o}) = \sigma_{xy}^{\rm H}(\mu,T)
 \label{eq:optimaldef}
\end{equation}
implies that
\begin{equation}
    -e^2 \sum_q\bm a_q^{\rm o}\cdot\bm\Xi_q(\mu,T) = \sigma_{xy}^{\rm H}(\mu,T) - \sigma_{xy}^{\rm B}(\mu,T; \bm{0}).
 \label{eq:optimaleq}
\end{equation}

A solution exists when at least one $\bm\Xi_q$ is nonzero, but it is generally nonunique and dependent on the chemical potential $\mu$ and the temperature $T$ and, as we discuss for a specific example in Sec.\ \ref{sec:numerics}, on details of the disorder distribution (even in the limit of vanishing disorder strength). Requiring the same ``optimal'' embedding at every $\mu$ and $T$ imposes a continuum of constraints, which generically do not have a solution. If the condition of embedding independence of Eq.~\eqref{eq:classcriterion} holds while $\sigma_{xy}^{\rm B}\neq\sigma_{xy}^{\rm H}$, changing the embedding cannot make the two conductivities equal. We thus conclude that, although an ``optimal'' embedding $\bm{a}^{\rm o}$ with $\sigma^{\rm B}_{xy}(\bm{a}^{\rm o}) = \sigma_{xy}^{\rm H}$ generically exists for given values of $\mu$ and $T$, it is not a useful concept, since one is required to first calculate the full Hall conductivity $\sigma_{xy}^{\rm H}$ for each $\mu$, $T$ in order to determine $\bm{a}^{\rm o}$. 

\section{embedding independence of the Hall conductivity: Case study for a graph with uncorrelated disorder}
\label{sec:disorder}

\begin{figure*}[t]
\includegraphics[width=.8\textwidth]{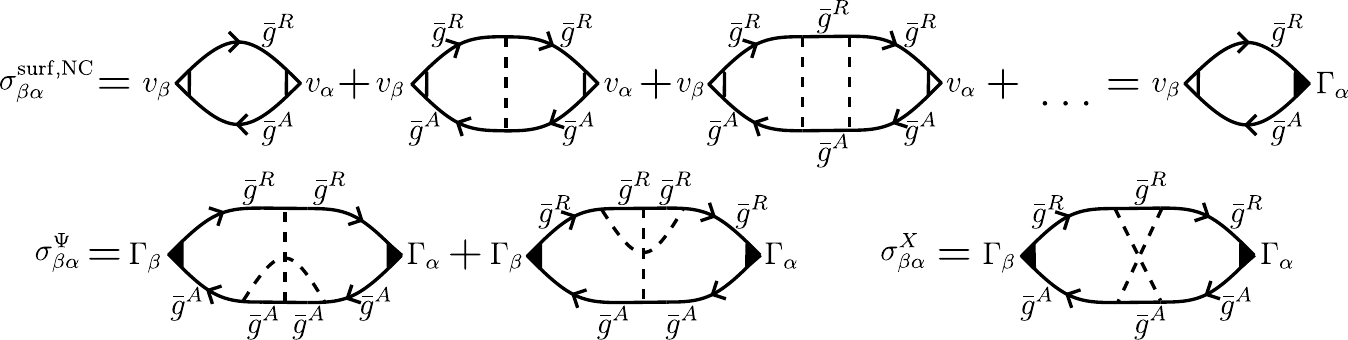}

\caption{Diagrams for the calculation of the Fermi-surface contribution to the Hall conductivity of a disordered periodic tight-binding graph. The lines with arrows and the open triangles represent disorder-averaged Green functions and velocity vertices, respectively. The top row shows the non-crossing diagrams and defines the vertex correction (solid triangle). The bottom row shows the crossed diagrams.
 \label{fig:diagrams}}
\end{figure*}

\subsection{Graph with uncorrelated disorder}

In Sec.\ \ref{sec:graph} we have presented general arguments why the Hall conductivity $\sigma_{xy}^{\rm H}$ is embedding-independent in the presence of disorder in the relaxation-dominated limit $\omega \tau \ll 1$. We now consider the specific case of a graph with uncorrelated disorder and show how the embedding independence is realized in a calculation of $\sigma_{xy}^{\rm H}$ for weak disorder using diagrammatic perturbation theory.

We consider a Hamiltonian of the form of Eq.\ (\ref{eq:graphH}) and include a Gaussian-distributed random on-site potential $u_{\bm{s}}$ with variance
\begin{equation}
  \langle u_{\bm{s}} u_{\bm{s'}} \rangle_{\rm dis} 
  - \langle u_{\bm{s}} \rangle_{\rm dis} \langle u_{\bm{s'}} \rangle_{\rm dis}
  = \gamma_q \delta_{\bm{s}\bm{s'}},
 \label{eq:diagdisorder}
\end{equation}
where we recall that $\bm{s} = (\bm{n},q)$ is the site index, see Sec.\ \ref{sec:graph}. Upon taking the disorder average, the conductivity can be arranged into the Kubo--St\v{r}eda form \cite{Bastin1971,Streda1982,Nagaosa2010},
\begin{equation}
    \sigma_{\beta\alpha}^H
    = \sigma_{\beta\alpha}^{\rm surf}
    + \sigma_{\beta\alpha}^{\rm sea},
    \label{eq:KSdecomp}
\end{equation}
where, in the dc limit,
\begin{align}
    \sigma_{\beta\alpha}^{\rm surf}
    &= \frac{e^2\hbar}{2\pi}\int dE\,
    \left(-\frac{df(E)}{dE}\right)
    \Tr \left\langle
    v_\beta G^R v_\alpha G^A
    \right\rangle_{\rm dis},
    \label{eq:KSsurf}
\end{align}
represents a contribution localized around the Fermi surface, while
\begin{align}
    \label{eq:KSsea}
    \sigma_{\beta\alpha}^{\rm sea} =&\, -\frac{e^2\hbar}{2\pi} \int dE\, f(E)\, \\ \nonumber &\, \ \ \ \ \mbox{} \times\Tr \left\langle v_\beta \partial_E G^R v_\alpha G^R  
     + v_\beta G^A v_\alpha \partial_E G^A \right\rangle_{\rm dis},
\end{align}
corresponds to the contribution from the full Fermi sea.\footnote{In Refs.\ \cite{Bastin1971,Streda1982,Nagaosa2010}, $\sigma^{\rm surf}$ is denoted $\sigma^{I(a)}$, whereas a partial integration is performed on $\sigma^{\rm sea}$, which turns it into the sum of the Fermi-surface contribution $\sigma^{I(b)}$ and the Fermi-sea contribution $\sigma^{II}$.} Here $G^{R/A}=(E-H\pm i \eta)^{-1}$ are the Green functions of the full disordered Hamiltonian, and $\Tr$ includes momentum integration and the trace over orbital indices, which we will denote with ``$\tr$". The Hall conductivity is obtained by antisymmetrizing in $\alpha$ and $\beta$.

The disorder average can be performed using standard methods of diagrammatic perturbation theory. The diagrams that have to be calculated in the limit of weak disorder (relaxation rate $\tau^{-1} \ll E_{\rm F}/\hbar$, where $E_{\rm F}$ is the Fermi energy) are shown in Fig.\ \ref{fig:diagrams}. These are the non-crossing diagrams (top row in Fig.\ \ref{fig:diagrams}) as well as selected diagrams with a single pair of crossing impurity lines \cite{Ado2015} (bottom row in Fig.\ \ref{fig:diagrams}). In App.\ \ref{app:crossed} we show that the antisymmetric part of the diagrams with the crossed impurity lines vanish for the case of weak uncorrelated disorder. Hence, it is sufficient to consider the non-crossing diagrams of the top row of Fig.\ \ref{fig:diagrams}.

The building block of the diagrammatic calculation is the disorder-averaged Green function
\begin{equation}
    \bG^{R/A}(\bm k,E) = \bigl[E-h_0(\bm k)-\Sigma^{R/A}(E)\bigr]^{-1}. \label{eq:GbarMain}
\end{equation}
The self energy $\Sigma^{R/A}(E)$ is calculated in the self-consistent Born approximation, \begin{equation}
    \Sigma^{R/A}(E) = \sum_q\gamma_qP_q\intk\bG^{R/A}(\bm k,E)P_q,
 \label{eq:scba}
\end{equation}
where $P_q = \ket{q}\bra{q}$ in the projector operator onto the $q$-th orbital of the unit cell. Because the correlator in Eq.~\eqref{eq:diagdisorder} is local and diagonal in $q$, the self-energy is momentum independent and diagonal in the orbital basis.

In the weak-disorder limit, the disorder average of the products of two retarded or two advanced Green functions in Eq.~\eqref{eq:KSsea} can be calculated as the product of the disorder-averaged Green functions. For the disorder average of the product of retarded and advanced Green functions in Eq.~\eqref{eq:KSsurf}, it is necessary to sum the full diagrammatic series of the top row of Fig.\ \ref{fig:diagrams}. Summation of the series amounts to replacing the velocity $v_{\alpha}$ by the vertex correction $\Gamma_{\alpha}$, so that the complete expression for $\sigma_{\beta\alpha}^{\rm surf}$ within the non-crossing approximation is
\begin{align}
    \sigma_{\beta\alpha}^{\rm surf, NC} =&\, \frac{e^2\hbar}{2\pi}\int dE\, \left( - \frac{df(E)}{dE} \right)
    \intk \notag
    \\ 
    & \mbox{} \times\tr\!\left[  v_\beta(\bm k)\bG^R(\bm k,E) \Gamma_\alpha(\bm{k}  ,E)\bG^A(\bm k,E) \right].
 \label{eq:fullRAbubble}
\end{align}
The full vertex $\bm{\Gamma}$ contains the bare velocity $\bm{v}$ and all ladder corrections. It obeys the Bethe--Salpeter equation
\begin{align}
 \label{eq:BSmain}
   \bm{\Gamma}(\bm{k},E) =&\, \bm{v}_{\bm{k}} + 
    \sum_q\gamma_q \intkp  
    \\ \nonumber &\, \mbox{} \times
    P_q \bG^R(\bm k',E) 
    \bm{\Gamma}(\bm k',E)\bG^A(\bm k',E)P_q.
\end{align}
The second term in Eq.\ (\ref{eq:BSmain}) is diagonal in the orbital basis and independent of $\bm{k}$. This makes an efficient numerical calculation of the vertex correction possible, as we discuss in Sec.\ \ref{sec:numerics}.

To connect these expressions to the semiclassical theory of the AHE, one projects the velocity vertices $v_{\alpha}$ and $v_{\beta}$ onto the band basis and decompose them into intraband and interband parts, Eqs.~\eqref{eq:vintra} and~\eqref{eq:vinter}.
As shown by Sinitsyn \textit{et al.}~\cite{SinitsynEtAl2007}, retaining the interband components of both $v_{\alpha}$ and $v_{\beta}$ yields the band contribution $\sigma^{\rm B}_{xy}$ of Eq.~\eqref{eq:berryformula}, whereas retaining the intraband components of both vertices gives the skew-scattering contribution. The side-jump and anomalous-distribution contributions arise from mixed combinations, with one velocity vertex taken to be interband and the other intraband. For a Gaussian random on-site potential, the skew-scattering contribution originates from scattering off pairs of nearby impurities, which is why we refer to it as ``pair-skew scattering.'' In the weak-disorder limit, the pair-skew-scattering contribution scales as $\tau^0$, like the other three contributions to the anomalous Hall conductivity. Non-Gaussian disorder can, in principle, also generate a skew-scattering contribution proportional to $\tau$~\cite{SinitsynEtAl2007,Sinitsyn2008,Nagaosa2010}. However, this contribution is absent in the Gaussian-disorder model considered here. The side-jump and anomalous-distribution contributions differ by which of the two velocity vertices is taken to be the interband one, which mirrors their different interpretations in the semiclassical framework \cite{Sinitsyn2008,Nagaosa2010}. Their contributions to the anomalous Hall conductivity are equal \cite{SinitsynEtAl2007}. 

The decomposition of $\sigma_{xy}^{\rm H}$ into these four semiclassical contributions does not coincide term by term with either its diagrammatic separation into the bare bubble, represented by the leftmost diagram in the top row of Fig.~\ref{fig:diagrams}, and the ladder corrections, represented by the remaining diagrams in that row, or the separation into the Fermi-sea and Fermi-surface terms of Eqs.~\eqref{eq:KSsurf} and~\eqref{eq:KSsea}. The band contribution arises entirely from the bare bubble, combining Eqs.~\eqref{eq:KSsurf} and~\eqref{eq:KSsea}, whereas the skew-scattering contribution arises from the ladder series.
 The side-jump and anomalous-distribution terms receive contributions from both the bare bubble and the ladder diagrams of the Fermi-surface contribution. Explicit expressions for each of the semiclassical contribution for the model with uncorrelated disorder are derived in App.~\ref{app:fullhall}.

\subsection{Embedding invariance of $\sigma_{xy}^{\rm H}$}

We compare the diagrammatic calculation of the Hall conductivity for two embeddings related by a displacement $\bm{a}_q$ of the lattice sites, see Eq.~\eqref{eq:shift}. We demonstrate that the individual contributions $\sigma^{\rm surf}$ and $\sigma^{\rm sea}$ of the Kubo--St\v{r}eda conductivity are not, by themselves, embedding-independent. The invariant object is the complete conserving response $\sigma^{\rm surf} + \sigma^{\rm sea}$. We first determine how $\sigma^{\rm surf}$ changes under a change of embedding and then show that the full Hall conductivity, $\sigma_{xy}^{\rm H}=\sigma^{\rm surf}_{xy}+\sigma^{\rm sea}_{xy}$, is embedding-independent.

Changing the embedding modifies the Bloch basis used to define the Bloch Hamiltonian $h_0(\bm{k})$, see Eq.~\eqref{eq:hchange}.
The disorder-averaged Green functions transform covariantly under the same basis transformation,
\begin{align}
    \widetilde\bG{}^{R/A}(\bm k,E) = U_{\bm k}\bG^{R/A}(\bm k,E)U_{\bm k}^{\dagger},
\end{align}
with $U_{\bm{k}}$ defined in Eq.~\eqref{eq:Udef}. 
For uncorrelated disorder, the self energy $\Sigma^{R/A}(E)$ is diagonal in the site basis and, hence, commutes with $U_{\bm{k}}$.

The velocity vertex does not transform purely covariantly because it contains a momentum derivative. Indeed,
\begin{align}
  \label{eq:valphatilde}
    \widetilde {\bm{v}}_{\bm{k}} &= \frac{1}{\hbar}  
    \bm{\nabla}_{\bm k} \left[U_{\bm k}h_0(\bm k)U_{\bm k}^{\dagger}\right] \notag\\
    &= U_{\bm k}\left( \bm{v}_{\bm{k}}-\frac{i}{\hbar}[\bm{a},h_0(\bm k)]\right) U_{\bm k}^{\dagger},
\end{align}
where $\bm{a} = \sum_{q} \bm{a}_q P_q$, see Eq.\ (\ref{eq:adef}). The commutator $[\bm{a},h_0({\bm k})]$ in the second line Eq.\ (\ref{eq:valphatilde}) mirrors the commutator in the embedding shift (\ref{eq:jtransform}) of the current density $\bm{j}$ in the second-quantized formulation of Sec.\ \ref{sec:graph}. The same commutator also appears in Refs.\ \cite{Boykin2001,Esteve-Paredes2023} when relating velocity operators in different embeddings.

We define $\delta_{\rm emb} \bm{v}$ as the residual change of the velocity after the common covariant rotation has been removed,
\begin{equation}
    \widetilde {\bm{v}}_{\bm{k}}=U_{\bm k}\left[ \bm{v}_{\bm{k}}+\delta_{\rm emb} \bm{v}_{\bm{k}}\right]U_{\bm k}^{\dagger}.
\end{equation}
Similarly, the change of the renormalized vertex under a change of embedding may be written as
\begin{equation}
    \widetilde{\bm{\Gamma}} =U_{\bm k}\left[\bm{\Gamma} + \delta_{\rm emb} \bm{\Gamma} \right]U_{\bm k}^{\dagger}.
\end{equation}
Calculating the conductivity contributions using Eqs.~\eqref{eq:KSsea} (with $G^{R/A}$ replaced by $\bG^{R/A}$) and~\eqref{eq:fullRAbubble}, we then see that all factors $U_{\bm k}^{\dagger}$ and $U_{\bm k}$ drop out of the calculation because of the cyclicity of the trace.
To determine the change in the Hall conductivity upon varying the embedding, it therefore remains to account for the changes $\delta_{\rm emb} \bm{v}$ and $\delta_{\rm emb} \bm{\Gamma}$ of the (renormalized) current vertices.

Upon inserting Eq.\ (\ref{eq:valphatilde}) for $\widetilde {\bm{v}}_{\bm{k}}$, we obtain that (cf. Eq.~\eqref{eq:jtransform})
\begin{equation}
  \label{eq:dv}
    \delta_{\rm emb} \bm{v}_{\bm{k}} = -\frac{i}{\hbar}[\bm{a},h_0(\bm{k})].
\end{equation}
Employing Eq.~\eqref{eq:GbarMain} to express $h_0(\bm{k})$ in terms of the inverse of the disorder-averaged Green function $\bG^{R/A}(\bm{k},E)$ and using that the self-energy $\Sigma^{R/A}(E)$ commutes with $a_{\alpha}$, we rewrite this as
\begin{equation}
   \delta_{\rm emb} \bm{v}_{\bm{k}}
    = \frac{i}{\hbar}[ \bm{a},\bG^{R/A}(\bm{k},E)^{-1}].
    \label{eq:embvelocitymain}
\end{equation}
The change $\delta_{\rm emb} \bm{\Gamma}(\bm{k},E)$ can be obtained from the Bethe--Salpeter equation~\eqref{eq:BSmain} with $\bm{v}(\bm{k})$ on the right-hand side replaced by $\delta_{\rm emb} \bm{v}$. Using the definition~\eqref{eq:scba} of the self energy, we obtain the retarded--advanced Ward identity
\begin{equation}
    \delta_{\rm emb} \bm{\Gamma}(\bm{k},E) = -\frac{i}{\hbar}\left[\bG^R(\bm{k},E)^{-1} \bm{a}-\bm{a}\,\bG^A(\bm{k},E)^{-1}\right]. \label{eq:dressedvertexWard}
\end{equation}

To make the cancellation among the Kubo--St\v{r}eda terms explicit, we isolate one external vertex at a time.
For simplicity, we consider the effect of the embedding transformation on the field-coupling vertex $\Gamma_\alpha$ in Eq.~\eqref{eq:fullRAbubble} and $v_\alpha$ in Eq.~\eqref{eq:KSsea}, while keeping the measured-current vertex $v_\beta$ unchanged. Substituting Eq.~\eqref{eq:dressedvertexWard} into Eq.~\eqref{eq:fullRAbubble}, we find that the embedding variation of the Fermi-surface term reduces to
\begin{equation}
    \delta_{\rm emb}^{(\alpha)} \sigma_{\beta\alpha}^{\rm surf, NC}
  = -\mathcal R_{\beta\alpha}^{(\alpha)}, \label{eq:laddervariation}
\end{equation}
where
\begin{align}
 \label{eq:Rexplicit}
    \mathcal R_{\beta\alpha}^{(\alpha)} & =\frac{i e^2}{2\pi} \int dE\,\left(-\frac{df(E)}{dE}\right) \\ \notag 
    & \quad\times\intk\tr\!\left[v_\beta (a_\alpha\bG^A(\bm{k},E)-\bG^R(\bm{k},E) a_\alpha) \right].
\end{align}
Thus, although Eq.~\eqref{eq:fullRAbubble} contains the full ladder series and a product of retarded and advanced Green functions, its embedding variation collapses to a single term linear in the Green functions and involving only the bare velocity $v_\beta$.

The Fermi-sea Kubo--St\v{r}eda contribution $\sigma^{\rm sea}_{\beta\alpha}$ acquires an embedding variation that is equal in magnitude and opposite in sign to that of the Fermi-surface term. This can be seen directly by substituting Eq.~\eqref{eq:embvelocitymain} into Eq.~\eqref{eq:KSsea}. For the first and second terms in Eq.~\eqref{eq:KSsea}, we express $\delta_{\rm emb}v_\alpha(\bm{k})$ in terms of $\bG^{R}(\bm{k},E)^{-1}$ and $\bG^{A}(\bm{k},E)^{-1}$, respectively. After straightforward algebra, we obtain
\begin{align}
\label{eq:seavariation}
    \delta_{\rm emb}^{(\alpha)} \sigma_{\beta\alpha}^{{\rm sea},{\rm NC}} =&\, \frac{i e^2}{2\pi} \int dE\, f(E) 
    \intk 
    \\ \nonumber &\, \mbox{} \times
    \tr \left[v_\beta (a_\alpha \partial_E \bG^A(\bm{k},E) - \partial_E \bG^R(\bm{k},E) a_\alpha )   \right]. 
\end{align}
In deriving this result, one also encounters terms of the form $v_\beta \partial_E \bG^R (\bG^R)^{-1} a_\alpha \bG^R$ and $v_\beta \bG^A  a_\alpha (\bG^A)^{-1} \partial_E \bG^A$. These terms can first be simplified using $\partial_E\bG^{R/A}= - \bG^{R/A}(1-\partial_E\Sigma^{R/A})\bG^{R/A}$. Their momentum integrals then vanish. Indeed, because the self-energy~\eqref{eq:scba} is momentum independent, this implies
\begin{equation}
    \bG^{R/A}(\bm{k},E) v_\beta\bG^{R/A}(\bm{k},E) = \frac{1}{\hbar} \partial_{k_\beta}\bG^{R/A}(\bm{k},E).
\end{equation}
Using this identity and the cyclicity of the trace, the corresponding momentum-space contributions can be written as total derivatives over the Brillouin-zone torus and therefore vanish by periodicity. Finally, integrating~\eqref{eq:seavariation} by parts, we verify that 
\begin{equation}
  \delta_{\rm emb}^{(\alpha)} \sigma_{\beta\alpha}^{{\rm sea},{\rm NC}}
  = +\mathcal R_{\beta\alpha}^{(\alpha)}.  \label{eq:remainingvariation}
\end{equation}
Together Eqs.~\eqref{eq:laddervariation} and~\eqref{eq:remainingvariation} therefore give
\begin{equation}
    \delta_{\rm emb}^{(\alpha)} \sigma_{\beta\alpha}^{\rm NC}(0) =0. \label{eq:currentvertexcancel}
\end{equation}
Similar calculations can be carried out for a change in the measured-current vertex, $\delta_{\rm emb}v_{\beta}$, or when both vertices, $\alpha$ and $\beta$, are varied simultaneously.

We have thus verified that within the non-crossing approximation the Hall conductivity $\sigma_{xy}^{\rm H}$ of a graph with uncorrelated on-site disorder is embedding invariant, whereas the individual contributions $\sigma^{\rm surf}_{xy}$ and $\sigma^{\rm sea}_{xy}$ are not. The embedding dependence of the individual contributions to the Hall conductivity can be traced back to the embedding dependence of the velocity vertex $\bm{v}_{\bm{k}}$, see Eq.\ (\ref{eq:dv}). Writing $\bm{v}_{\bm{k}}$ as a sum of interband and intraband contributions, $\bm{v} = \bm{v}^{\rm intra} + \bm{v}^{\rm inter}$, see Eqs.\ (\ref{eq:vintra}) and (\ref{eq:vinter}), we see that it is only the interband term that depends on the choice of the embedding, 
\begin{equation}
    \delta_{\rm emb} \bm{v}^{\rm intra}_{\bm k} = 0.
    \label{eq:dvembintra}
\end{equation}
Of the four contributions to $\sigma_{xy}^{\rm H}$ in the semiclassical theory of the AHE, the skew-scattering term has two intraband velocity vertices, so that it is embedding-independent. The remaining three contributions, the band, side-jump, and anomalous-distribution terms, contain at least one interband velocity vertex, which makes them embedding-dependent. 

We did not need to consider the crossed diagrams, because they vanish for a model with uncorrelated disorder, see App.~\ref{app:crossed}. Crossed diagrams typically give a contribution to the Hall conductivity that is comparable to that of the non-crossing diagrams in the general case of non-diagonal, correlated, or non-point-like disorder \cite{Ado2015}. Crossed diagrams are analogous to the pair-skew-scattering contribution in that they involve two intraband velocity vertices. Since the intraband velocity vertex is embedding covariant, see Eq.\ (\ref{eq:dvembintra}), the contribution to the Hall conductivity from crossed diagrams is embedding-independent, even if these diagrams are nonzero.

\section{Numerical illustration}
\label{sec:numerics}

We test the embedding invariance of the complete non-crossing Hall conductivity  for two concrete lattice models. As our first example, we consider a two-band model, for which the orbital positions are shifted continuously within a fixed unit cell. As a second example, in App.\ \ref{app:numerics} we discuss a flux-threaded kagome lattice. There, we consider an embedding in which all orbitals in the same unit cell are assigned the same position (the so-called ``periodic gauge'') for two different choices of the unit cell.

\subsection{Numerical procedure}
\label{sec:num_method}

For each embedding, we solve the self-consistent Born equation~\eqref{eq:scba} for the self-energy $\Sigma^{R/A}(E)$ and the Bethe--Salpeter equation~\eqref{eq:BSmain} for the vertex correction $\bm{\Gamma}$ in the non-crossing approximation. The calculation of $\Sigma^{R/A}(E)$ and $\bm{\Gamma}(\bm{k},E)$ is numerically efficient, because $\Sigma^{R/A}(E)$ and $\delta \bm{\Gamma}(E) = \bm{\Gamma}(\bm{k},E)-\bm{v}_{\bm{k}}$ do not depend on $\bm{k}$ for a model with uncorrelated on-site disorder. We then evaluate the complete Kubo response $\sigma_{xy}^{\rm H}$ from Eqs.\ \eqref{eq:KSsea} (with $G^{R/A}$ replaced by $\bG^{R/A}$) and~\eqref{eq:fullRAbubble}. We obtain the four contributions $\sigma_{xy}^{X}$ from the semiclassical theory, with $X = $``B'', ``s.j.'', ``a.d.'', ``p.s.'' standing for the band, side jump, anomalous distribution, and pair-skew scattering terms, by repeating this procedure with the velocity vertices $v_{\alpha}$ and $v_{\beta}$ projected onto their intra- and inter-band contributions, as described in Sec.\ \ref{sec:disorder}. Our numerical calculations allow us to directly verify that the full Kubo response $\sigma_{xy}^{\rm H} = \sigma_{xy}^{\rm B} + \sigma_{xy}^{\rm s.j.} + \sigma_{xy}^{\rm a.d.} + \sigma_{xy}^{\rm p.s.}$ is embedding-independent.

The Brillouin-zone integrals are evaluated on a uniform momentum grid and calculation of the self-energy is iterated until convergence at every energy $E$ separately. We verify convergence with respect to the momentum mesh, energy resolution, and self-consistency tolerance. The disorder strength is taken small enough, that the Hall conductivity no longer depends on it. More details about our numerical evaluations can be found in App.~\ref{app:numerics}. 

\subsection{Two-band Chern insulator} \label{subsec:twoband}

\begin{figure*}[t!]
    \centering
    \includegraphics[width=\textwidth]{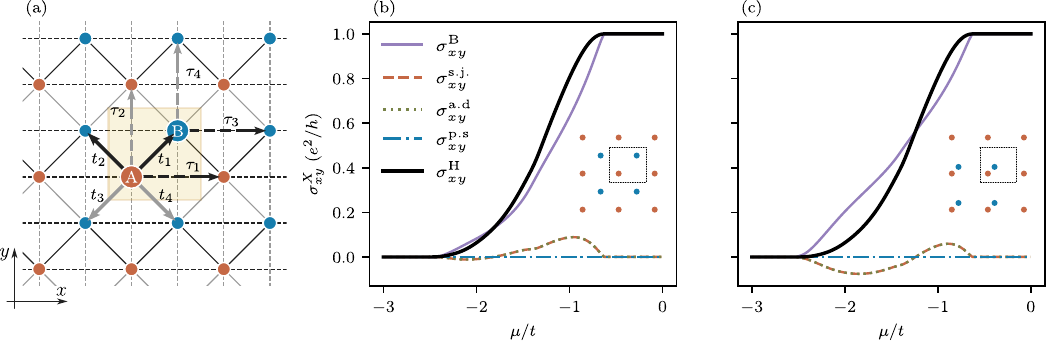}
    \caption{Embedding dependence of the Hall conductivity for the two-band Chern insulator model of Eq.\ (\ref{eq:nnn-hamiltonian}) with $\gamma_A = \gamma_B = 0.05 t^2$ for $\zeta=0.5$. (a) Real-space lattice for $\bm{\rho}_{A} = (0,0)$, $\bm{\rho}_{B} = (0.5,0.5)$ and the two-site unit cell (shaded). Black bonds are independent of the anisotropy parameter $\zeta$, whereas gray bonds are proportional to it, see Eq.\ (\ref{eq:numerical-hoppings}). (b,c) Antisymmetrized Hall-conductivity $\sigma_{xy}^{\rm H}$ and its four contributions $\sigma_{xy}^{\rm B}$, $\sigma_{xy}^{\rm s.j.}$, $\sigma_{xy}^{\rm a.d.}$, and $\sigma_{xy}^{\rm p.s.}$ versus chemical potential $\mu$ for the embeddings $\bm{\rho}_{A} = (0,0)$, $\bm{\rho}_{B} = (0.1,0.1)$ and $\bm{\rho}_{A} = (0,0)$, $\bm{\rho}_{B} = (0.5,0.5)$, respectively, which correspond to the two lattices shown in Fig.\ \ref{fig:graph_embedding}. If the chemical potential is in the band gap, $-0.67 \lesssim \mu/t \lesssim 0.67$, the Hall conductivity is given by the band contribution $\sigma_{xy}^{\rm B}$ only, which is quantized and embedding-independent. If $\mu$ is inside a band, $\sigma_{xy}^{\mathrm{H}}$ and $\sigma_{xy}^{\mathrm{p.s.}}$ are insensitive to changes in the embedding, whereas the remaining contributions are not. The side-jump and anomalous distribution terms are equal, as evident from the middle and right panels.
  }
    \label{fig:four-bond-embedding}
\end{figure*}

When considering a two-band system, the general formulas simplify substantially. For
\begin{equation}
    h_0(\bm{k})=d_0(\bm{k})\sigma_0+\bm d(\bm{k})\cdot\bm\sigma,
 \qquad d=|\bm d|,
 \label{eq:twobandH}
\end{equation}
with eigenvalues $\varepsilon_s=d_0+sd$ and $s=\pm$, the Berry curvature is
\begin{align}
    \Omega_{s,z}&=-\frac{s}{2}\hat{\bm d}\cdot \left(\partial_{k_x}\hat{\bm d}\times\partial_{k_y}\hat{\bm d}\right),
 \label{eq:twobandberry}
\end{align}
where $\hat{\bm{d}} = \bm{d}/d$ is a unit-vector. Identifying the two orbitals $A$ and $B$ with the $\sigma_z$ basis: $P_A =(\sigma_0+\sigma_z)/2$ and $P_B=(\sigma_0-\sigma_z)/2$, the corresponding orbital weights are
\begin{equation}
 w_{s,A/B}=\frac{1}{2}(1\pm s\hat d_z).
 \label{eq:twobandweights}
\end{equation}
With this notation, the relevant energy-resolved vector that measures embedding sensitivity reads
\begin{align}
 \bm\xi_A(E)&=-\bm\xi_B(E)
 =\frac{1}{2\hbar}\sum_{s=\pm}s
 \oint_{\varepsilon_s=E}
 \frac{\hat d_z(\bm k)\,d\bm k}{(2\pi)^2}.
 \label{eq:twobandxi}
\end{align}
Thus the clean embedding problem is controlled by one relative displacement $\bm{a}_A - \bm{a}_B$ and one contour vector $\bm\xi_A(E)$. In the remainder of this section, we set $\bm a_A$ to 0 and only vary $\bm a_B$.

As a concrete example, we consider a two-band model defined on the graph shown in Fig.~\ref{fig:four-bond-embedding}, with sublattices $A$ and $B$, nearest- and next-nearest-neighbor hoppings, a staggered on-site potential, and uncorrelated on-site disorder. We set the lattice constant to 1 and measure all momenta in the corresponding dimensionless units. 

The Hamiltonian of this model reads
\begin{equation}
h_0(\bm{k})
=
\begin{pmatrix}
h_A(\bm{k}) + u & f^*(\bm{k})\\
f(\bm{k})   & h_B(\bm{k}) - u 
\end{pmatrix},
\label{eq:nnn-hamiltonian}
\end{equation}
where the inter-sublattice contribution from nearest-neighbor hopping is
\begin{align}
    f(\bm{k}) = &\, e^{-i\bm{k}\cdot(\bm{\rho}_B-\bm{\rho}_A)} \left[ t_1+t_2e^{ik_x} +t_3e^{i(k_x+k_y)} +t_4e^{ik_y} \right], \nonumber \\
    h_A(\bm{k}) =&\, \tau_1 e^{ik_x}+\tau_2 e^{ik_y}+\mathrm{c.c.},
    \nonumber \\
    h_B(\bm{k}) =&\, \tau_3 e^{ik_x} + \tau_4 e^{ik_y} +\mathrm{c.c.}
\label{eq:nn-hopping}
\end{align}
Here $t_1,\ldots,t_4$ and $\tau_1,\ldots,\tau_4$ are the four nearest-neighbor and next-nearest-neighbor hopping
amplitudes, respectively, as indicated in Fig.\ \ref{fig:four-bond-embedding}(a), and $\bm{\rho}_A$ and $\bm{\rho}_B$ are the embedding vectors of the graph, see Eq.\ (\ref{eq:embedding}).
For simplicity, we restrict our choice of model parameters for $h_0(\bm{k})$ to a one-parameter set labeled by an ``anisotropy parameter'' $\zeta$,
\begin{equation}
\begin{aligned}
    u & = -\frac{t}{2}, \\ 
    (t_1,t_2,t_3,t_4) & = t\left(-i,\, 1,\, i \zeta,\, -\zeta \right), \\
    (\tau_1,\tau_2,\tau_3,\tau_4) & = \frac{t}{2} \left( -1,\, -\zeta,\, 1,\, \zeta \right).
\end{aligned}
\label{eq:numerical-hoppings}
\end{equation}

In Fig.~\ref{fig:xi_concept}, we illustrate that the energy-resolved vector $\bm{\xi}_A(E)$ vanishes in the inversion-symmetric limit $\zeta=1$. On the other hand, for $\zeta = 0.5$, inversion symmetry is broken, $\xi_A(E)$ becomes finite and the band contribution to the Hall conductivity is embedding-dependent.

Figure \ref{fig:four-bond-embedding} shows numerical results for the Hall conductivity for $\zeta = 0.5$  and dimensionless disorder strength $\gamma_A = \gamma_B = \gamma$. For $\zeta = 0.5$, the model \eqref{eq:nnn-hamiltonian} describes a Chern insulator with a band gap for $-0.67 \lesssim \mu/t \lesssim 0.67$. If the chemical potential $\mu$ is in the band gap,
the Hall conductivity is given by the band contribution $\sigma_{xy}^{\rm B}$, which is quantized and embedding-independent. If the chemical potential is inside the bands, the full Hall conductivity $\sigma_{xy}^{\rm H}$ is still embedding-independent, but the contributions $\sigma_{xy}^{\rm B}$, $\sigma_{xy}^{\rm s.j.}$, and $\sigma_{xy}^{\rm a.d.}$ separately are not. In Fig.\ \ref{fig:four-bond-embedding} we illustrate this by showing all four contributions to $\sigma_{xy}^{\rm H}$ as a function of $\mu/t$ for the two different choices of the embedding shown in Fig.\ \ref{fig:graph_embedding} (b,c). (Note that the side-jump and anomalous-distribution contributions are equal and, hence, appear as a single curve in Fig.\ \ref{fig:graph_embedding}.)
The pair-skew contribution $\sigma_{xy}^{\rm p.s.}$ is embedding-independent, consistent with its semiclassical interpretation in terms of two intraband velocity vertices, whose diagonal matrix elements are determined solely by the band dispersion and are therefore unchanged under a change of embedding.

\begin{figure}[t!]
    \centering    \includegraphics[width=0.95 \columnwidth]{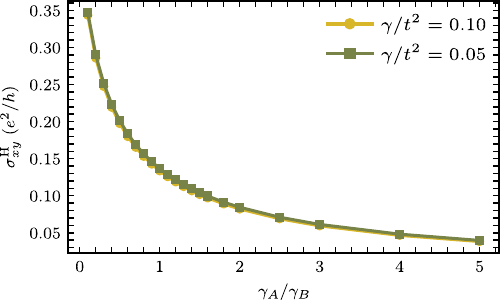}
    \caption{Disorder-ratio dependence of the total Hall conductivity using the two-band Chern insulator model at $\zeta = 0.5$ and $\mu/t=-1.8$. Results for $\gamma/t^2=0.10$ and $0.05$ are overlaid as a convergence check while varying
    the relative disorder strength $\gamma_A/\gamma_B$. }
    \label{fig:disorder-ratio}
\end{figure}

Although the Hall conductivity $\sigma_{xy}^{\rm H}$ is independent of the over-all strength $\gamma = (\gamma_A+\gamma_B)/2$ of the on-site disorder in the weak-disorder limit $
\gamma/t^2 \ll 1$, it does depend on the ratio $\gamma_A/\gamma_B$ of the disorder strengths on the two sublattices. This is illustrated in Fig.\ \ref{fig:disorder-ratio}, where we show $\sigma_{xy}^{\rm H}$ as a function of $\gamma_A/\gamma_B$ at a fixed value of $\gamma = (\gamma_A + \gamma_B)/2$ and for $\mu/t = -1.8$. The dependence on details of the disorder shown in Fig.\ \ref{fig:disorder-ratio} confirms a similar observation made in Ref.~\cite{Nagaosa2010}.

In the limit $\zeta = 0$, the graph described by the model (\ref{eq:nnn-hamiltonian}) separates into a collection of disconnected one-dimensional chains, as shown in the left panels of Fig.\ \ref{fig:cut-bond-embedding}.
Clearly, the Hall conductivity vanishes for such a system. However, depending on the embedding, this is not the case for the individual contributions. To illustrate this, in Fig.\ \ref{fig:cut-bond-embedding} we show these individual contributions to $\sigma_{xy}^{\rm H}$ for the embeddings $\bm\rho_A = \bm\rho_B = (0,0)$ (top panels) and $\bm\rho_A = (0,0)$, $\bm\rho_B = (0.5,0.5)$ (bottom panels). For the latter choice of the embeddings, the figure shows that band and side-jump/anomalous-distribution terms are nonzero, whereas the full Hall response $\sigma_{xy}^{\rm H}$ vanishes, as expected. For the former choice, all contributions to the Hall conductivity vanish, as shown in the upper panel of Fig.\ \ref{fig:cut-bond-embedding}.

\section{Discussion and conclusion}
\label{sec:discussion}

To obtain the band structure of a tight-binding model, it is sufficient to know the hopping amplitudes and on-site potentials; apart from the lattice periodicity, no information about the precise positions of the lattice sites within the unit cell is required. The same independence from the geometric embedding extends to long-distance wavefunction properties, such as the localization length in the presence of disorder. By contrast, observables such as the polarization and orbital magnetic moment manifestly depend on the orbital positions~\cite{FuchsPiechon2021,ThonhauserEtAl2005}. Accordingly, Wannier- and local-orbital-based calculations require accurate position matrix elements to describe electromagnetic response~\cite{WangYatesSouzaVanderbilt2006,Esteve-Paredes2023}, and geometric embedding have been optimized for the calculation of geometry-dependent quantities in a tight-binding framework \cite{TelleLange2026}. Within this dichotomy, the anomalous Hall conductivity $\sigma_{xy}^{\rm H}$ occupies an intermediate position. The microscopic expressions for the hopping amplitudes in the presence of an electric field and for the current density depend on the orbital embedding, yet the dc Hall conductivity itself does not, provided there is a minimal amount of relaxation ensuring that the current response to a time-independent electric field is itself time independent. In this article, we have established this result using general arguments and an explicit calculation of the anomalous Hall conductivity in a model with uncorrelated on-site disorder.

In the literature, the anomalous Hall conductivity is conventionally decomposed into four contributions: the ``intrinsic'' band contribution $\sigma_{xy}^{\rm B}$ and three ``extrinsic'', disorder-related contributions -- the side-jump, anomalous-distribution, and skew-scattering contributions. The first three of these are embedding-dependent, whereas the skew-scattering term and the full anomalous Hall conductivity $\sigma_{xy}^{\rm H}$ are not. The cancellation of the embedding dependencies of $\sigma_{xy}^{\rm B}$ and the side-jump and anomalous-distribution terms may seem surprising at first sight, because $\sigma_{xy}^{\rm B}$ is expressed as a Fermi-sea integral, whereas the other two contributions have support only at the Fermi surface~\cite{SimonRudner2020}. An explicit calculation of the embedding-dependent part of $\sigma_{xy}^{\rm B}$ shows, however, that this is itself entirely a Fermi-surface quantity, see Eq.~\eqref{eq:DeltasigmaXi}. It can therefore be canceled by the embedding-dependent parts of the disorder-related extrinsic Fermi-surface contributions without canceling the full band contribution itself. In the tight-binding model with uncorrelated on-site disorder, this cancellation is enforced by a Ward identity.

The presence of relaxation, {\em e.g.}, due to elastic scattering from an on-site disorder potential, ensures that the full conductivity tensor $\sigma_{\alpha\beta}$ is well defined at zero frequency. In the absence of relaxation, the symmetric part of the conductivity tensor diverges in the zero-frequency limit, corresponding to a current response that grows linearly in time under a time-independent applied electric field, whereas its antisymmetric part remains finite. In this case, the finite antisymmetric response coefficient depends on the geometric embedding for a generic tight-binding model. This embedding dependence reflects a sensitivity to microscopic details of how the current is measured (which cross section?) and how the electric field is applied (where is the potential drop?). Because of this, the Hall response in the absence of relaxation is not described by a unique bulk response coefficient and the notion of a ``Hall conductivity'' is not well defined. Instead, the current response of a ballistic system is described by a set of conductance coefficients \cite{Buettiker1996} that depend on the device geometry and size or, equivalently, by a non-local conductivity tensor \cite{Akkermans2010}.

For models with sufficient symmetry, the band contribution $\sigma_{xy}^{\rm B}$ to the Hall conductivity may become independent of the embedding, as discussed in Sec.\ \ref{sec:class}. This could suggest that such models possess a well-defined bulk anomalous Hall conductivity even in the absence of relaxation. However, this embedding independence is a fine-tuned property and need not survive in a transport setup with external contacts, since crystal boundaries and contacts generally break the symmetries responsible for it. Thus, while an embedding dependence demonstrates that $\sigma_{xy}^{\rm B}$ is not a measurable bulk property, symmetry-enforced embedding independence alone is not sufficient to establish that $\sigma_{xy}^{\rm B}$ provides a controlled approximation to the Hall conductivity measured in a device with external contacts.

Relaxation is not only essential for the existence of a well-defined, embedding-independent anomalous Hall conductivity $\sigma_{xy}^{\rm H}$, but it also leaves an imprint on its value, even in the weak-disorder limit, in which the Hall conductivity becomes independent of the overall disorder strength~\cite{Nagaosa2010}. In particular, scattering-induced contributions of order $\tau^0$, where $\tau$ is the mean relaxation time, remain finite in the weak-disorder limit. On the other hand, while relaxation renders the Hall conductivity $\sigma_{xy}^{\rm H}$ well defined, its value generally still depends on the details of the disorder and, in particular, on the relative disorder strengths on different orbitals as their common scale tends to zero, as we illustrate in Fig.~\ref{fig:disorder-ratio}. In this sense, $\sigma_{xy}^{\rm H}$ is a property of the tight-binding graph together with its relaxation mechanism, rather than of the clean Hamiltonian $H_0$ alone. Choosing more realistic orbital positions does not eliminate the need to calculate the complete response~\cite{WeischenbergEtAl2011}, including contributions arising from the relevant relaxation channels.

\begin{figure}
    \centering
    \includegraphics[width=0.95\columnwidth]{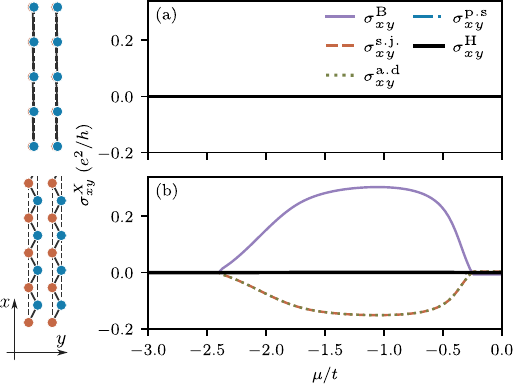}
    \caption{Hall conductivity of the model~\eqref{eq:nnn-hamiltonian} for $\zeta=0$. In this limit, the graph decomposes into disconnected one-dimensional chains (left panels). The complete Hall response vanishes, whereas the individual embedding-dependent contributions can remain finite, depending on the embedding. This is illustrated in the right panels for the choices $\bm\rho_A = \bm\rho_B = (0,0)$ (top) and $\bm\rho_A=(0,0)$, $\bm\rho_B = (0.5,0.5)$ (bottom).}
    \label{fig:cut-bond-embedding} 
\end{figure}

In some materials, the band contribution $\sigma_{xy}^{\rm B}$ provides a good quantitative estimate of the anomalous Hall conductivity despite neglecting relaxation-dependent contributions. One example is the agreement between $\sigma_{xy}^{\rm B}$ and the measured anomalous Hall conductivity in the ferromagnetic semiconductors (In,Mn)As and (Ga,Mn)As, reported in one of the original works proposing this contribution to the Hall conductivity~\cite{Jungwirth2002}. Other established examples include SrRuO${}_3$~\cite{Fang2003} or bcc Fe~\cite{Yao2004}. Such agreement, however, is not general. The tight-binding graph with uncorrelated on-site disorder considered here provides a simple extreme counterexample. Setting $\zeta=0$ in Eq.~\eqref{eq:numerical-hoppings} decomposes the graph into disconnected one-dimensional chains, for which the Hall conductivity $\sigma_{xy}^{\rm H}$ must vanish. Nevertheless, depending on the geometric embedding, the band contribution $\sigma_{xy}^{\rm B}$ remains finite, as shown in Fig.~\ref{fig:cut-bond-embedding}. Thus, although $\sigma_{xy}^{\rm B}$ may in some cases be close to the true Hall conductivity in the presence of relaxation, there is no {\em a priori} guarantee that it provides a reliable approximation.

In their study of embedding dependence in tight-binding models, Simon and Rudner emphasized embedding independence as a key requirement for expressions describing physical observables~\cite{SimonRudner2020}, alongside other fundamental invariance principles such as gauge invariance. Reference \cite{SimonRudner2020} applies this principle to the so-called ``geometric stability hypothesis'' for fractional Chern insulators~\cite{Jackson2015}, arguing that this criterion for maximizing the many-body spectral gap, which is itself embedding-independent, cannot be formulated solely in terms of the Berry curvature, which depends on the embedding. The present work shows that the anomalous Hall conductivity is likewise embedding-independent, even though its conventional decomposition into four contributions contains terms that individually depend on the embedding. Reference \cite{MorenoGonzalez2023} argues that the critical delocalized state near the center of a Chern band occurs at the energy $\varepsilon_{\rm cr}$ for which $h\sigma_{xy}^{\rm B}/e^2 = 1/2 \mod 1$, again relating an embedding-independent quantity to an embedding-dependent one. This relation can be made embedding-independent by replacing $\sigma_{xy}^{\rm B}$ with $\sigma_{xy}^{\rm H}$, consistent with Pruisken's identification of the critical values of $\sigma_{xy}^{\rm H}$ at the plateau transition of the quantized Hall effect~\cite{Pruisken1984}.

Apart from using embedding dependence as a check on physical relations, it is important to classify physical observables into embedding-dependent quantities, for which geometric data and microscopic choices are essential, and embedding-independent ones, which are independent of microscopic details, such as bond paths, local partitions of charge or current, or the microscopic distribution of the applied field. Related questions arise in thermal transport, where transport currents must be separated from circulating magnetization currents~\cite{QinNiuShi2011,ErcoleEtAl2016,HubnerEtAl2023}, and in spin transport, where spin nonconservation introduces torque terms~\cite{ShiEtAl2006}. Finite-frequency and finite-wave-vector response, correlated disorder, interactions, and inelastic relaxation provide further directions for determining which microscopic geometric sensitivities disappear in the stationary limit and which remain physically observable.

The broader lesson is that a contribution to a response function, a property of a particular spatial realization, and a stationary transport coefficient need not contain the same information. For the anomalous Hall effect studied here, relaxation and the complete transport response remove the dependence on orbital embedding, allowing for a significant amount of freedom when arranging the calculation of $\sigma_{xy}^{\rm H}$ for a specific model. Determining when analogous invariance emerges for other observables is a natural direction for future work.

\begin{acknowledgements}
We thank Alexander Altland, Adam Chaou, Lumen Eek, Leonid Glazman, Felix von Oppen, Steve Simon, and Ady Stern for helpful discussions. This work was supported by the Deutsche Forschungsgemeinschaft (DFG, German Research Foundation) as part of the German Excellence Strategy - EXC3112/1 - 533767171 (Center for Chiral Electronics) and CRC TR 183 (Project Number 277101999, subproject A02).

The numerical code used in this paper is openly available at \cite{zenodo_link}. We used OpenAI Codex (GPT-5.6 Sol) to assist in writing the code for the numerical calculations. The authors reviewed and tested the generated code and verified the numerical results. The authors take full responsibility for the calculations and conclusions presented in this work.
\end{acknowledgements}

\appendix

\section{Derivation of the complete leading weak-disorder Hall response}
\label{app:fullhall}

Throughout the Appendices, we use the thermodynamic-limit convention
\begin{equation}
    \frac{1}{L^2}\sum_{\bm k} \ldots
    \longrightarrow\int_{\bm k} \ldots
    \equiv\intk \ldots.
 \label{eq:appMomentumConvention}
\end{equation}
All momentum-space conductivity formulas are therefore written per unit area and contain no additional factor of $L^{-2}$. The symbol $\tr$ denotes the trace over orbital indices, whereas $\lambda$ and $\lambda'$ label bands. Clean and disorder-averaged Green functions are denoted by $g^{R/A}$ and $\bG^{R/A}$, respectively.

This first Appendix derives the band-resolved Hall conductivity through order $O[(\hbar/E_F\tau)^0]$. We first identify the conventional intrinsic term, including the off-shell retarded--advanced, retarded--retarded, advanced--advanced, and Fermi-sea sectors. We then add the transport-pole corrections generated by the ladder vertex and the band-off-diagonal self-energy.
Throughout this Appendix, band matrix elements are defined as
\begin{equation}
    (X_\alpha)_{\lambda\lambda'}\equiv\bra{u_{\lambda\bm k}}X_\alpha(\bm k)\ket{u_{\lambda'\bm k}},
 \label{eq:appBandMatrixElement}
\end{equation}
and the momentum argument is suppressed whenever no ambiguity can arise.

\subsection{Kubo--St\v{r}eda sectors and the intrinsic term}

The two Kubo--St\v{r}eda sectors are given in Eqs.~\eqref{eq:KSsurf} and \eqref{eq:KSsea}. The retarded--advanced term in Eq.~\eqref{eq:KSsurf} contains both intraband and interband parts. The band-diagonal Fermi-surface part gives the symmetric Drude response. The (intrinsic) band contribution to the Hall conductivity is instead obtained from
\begin{equation}
    \sigma_{\beta\alpha}^{\rm H,intr} \equiv \left[\sigma_{\beta\alpha}^{\rm surf}
    + \sigma_{\beta\alpha}^{\rm sea}\right]^{\rm H, diag},
\label{eq:appSeaOffDefinition}
\end{equation}
see Ref.~\cite{SinitsynEtAl2007}, where the superscript ``H'' denotes the part odd under $\alpha\leftrightarrow\beta$, while the superscript ``diag'' indicates that no ladder is attached to either external vertex and only the band-diagonal part of the self-energy is contained in the dressed Green's functions. Note that the band contribution is not obtained from the Fermi-surface part of the $RA$ bubble alone: the $RR$ and $AA$ sectors must be included as well.

This bookkeeping is particularly transparent in the clean limit. Define
\begin{equation}
\begin{aligned}
     \mathcal S_0(\bm k,E)&=\frac{i}{2\pi}\left[g^R(\bm k,E)-g^A(\bm k,E)\right]\\ 
     &=\sum_\lambda P_{\lambda\bm k}\delta(E-\varepsilon_{\lambda\bm k}),
\end{aligned}
 \label{eq:appCleanSpectral}
\end{equation}
where $P_{\lambda\bm k}=\ket{u_{\lambda\bm k}}\bra{u_{\lambda\bm k}}$. Combining the clean versions of \eqref{eq:KSsurf}-\eqref{eq:KSsea} and dropping a total momentum derivative gives
\begin{equation}
\begin{aligned}
    \sigma_{\beta\alpha}^{(0)} = & i e^2\hbar\int dE\,f(E) \notag \\
    &\quad \times \int_{\bm k}\tr\! \big[v_\beta\partial_E g^Rv_\alpha\mathcal S_0 - v_\beta\mathcal S_0v_\alpha\partial_E g^A\big].
 \label{eq:appCleanKubo}
 \end{aligned}
\end{equation}
Taking the spectral function on band $\lambda$ and both velocity operators off diagonal in the band basis yields
\begin{align}
    \sigma_{\beta\alpha}^{\rm B}&=-i e^2\hbar\sum_\lambda\int_{\bm k}f(\varepsilon_{\lambda\bm k}) \notag\\
    &\quad\times\sum_{\lambda'\neq\lambda}\frac{(v_\beta)_{\lambda\lambda'}(v_\alpha)_{\lambda'\lambda}-(v_\alpha)_{\lambda\lambda'}(v_\beta)_{\lambda'\lambda}}{(\varepsilon_{\lambda\bm k}-\varepsilon_{\lambda'\bm k})^2}. \label{eq:appIntrinsicVelocity}
\end{align}
For $\lambda' \neq \lambda$, differentiation of the eigenvalue equation gives
\begin{equation}
    (v_\alpha)_{\lambda\lambda'}=\frac{\varepsilon_{\lambda'\bm k}-\varepsilon_{\lambda\bm k}}{\hbar}\braket{u_{\lambda\bm k}}{\partial_{k_\alpha}u_{\lambda'\bm k}}.
 \label{eq:appOffDiagonalVelocity}
\end{equation}
Substituting Eq.~\eqref{eq:appOffDiagonalVelocity} into Eq.~\eqref{eq:appIntrinsicVelocity} gives
\begin{align}
    \sigma_{\beta\alpha}^{\rm B}&=-\frac{i e^2}{\hbar}\sum_\lambda\int_{\bm k}f(\varepsilon_{\lambda\bm k}) \notag\\
    &\quad\times\left[\braket{\partial_{k_\beta}u_{\lambda\bm k}}{\partial_{k_\alpha}u_{\lambda\bm k}}-\braket{\partial_{k_\alpha}u_{\lambda\bm k}}{\partial_{k_\beta}u_{\lambda\bm k}}\right],
 \label{eq:appIntrinsicBerry}
\end{align}
which is Eq.~\eqref{eq:berryformula}. In the weak-disorder expansion, self-energy corrections to the combination in Eq.~\eqref{eq:appSeaOffDefinition} that gives the band contribution are not enhanced by the Fermi-surface transport pole and first enter at relative order $\hbar/(E_F\tau)$. Consequently,
\begin{equation}
    \sigma_{\beta\alpha}^{\rm H,  intr}=\sigma_{\beta\alpha}^{\rm B}+O\!\left(\frac{e^2}{\hbar}\frac{\hbar}{E_F\tau}\right).
 \label{eq:appSeaOffBerry}
\end{equation}

\subsection{Fermi-surface corrections}

Following Ref.\ \cite{SinitsynEtAl2007}, the remaining extrinsic $O(\tau^0)$ Hall terms originate from the Fermi-surface part of the retarded--advanced sector in Eq.~\eqref{eq:KSsurf}. The non-crossing ladder series dresses the current vertex $\bm{\Gamma}(\bm{k}, E)$, which satisfies the Bethe-Salpeter equation \eqref{eq:BSmain}. The Fermi-surface projection of a retarded--advanced pair is
\begin{align}
    \bG^R(\bm k,E) \ldots \bG^A(\bm k,E)&\longrightarrow\sum_\lambda\frac{2\pi\tau_{\lambda\bm k}}{\hbar}\delta(E-\varepsilon_{\lambda\bm k}) \notag\\
    &\quad\times P_{\lambda\bm k} \ldots P_{\lambda\bm k}, \label{eq:appRAprojection}
\end{align}
with 
\begin{equation}
\begin{aligned}
    \eta(E)&=\frac{i}{2}\left[\Sigma^R(E)-\Sigma^A(E)\right],\\
    \eta_{\lambda\bm k}&=\bra{u_{\lambda\bm k}}\eta(\varepsilon_{\lambda\bm k})\ket{u_{\lambda\bm k}}=\frac{\hbar}{2\tau_{\lambda\bm k}}.
\end{aligned}
 \label{eq:appEtaDefinition}
\end{equation}
and
\begin{equation}
  P_{\bm{k}\lambda} = \ket{u_{\bm k\lambda}}\bra{u_{\bm k\lambda}}.
\end{equation}
When both external vertices are diagonal on the Fermi-surface band, Eq.~\eqref{eq:appRAprojection} gives the $O(\tau)$ Drude response, which is symmetric in $\alpha$ and $\beta$. The Hall response at $O(\tau^0)$ is obtained by retaining one Fermi-surface band $\lambda$ and one band $\lambda'\neq\lambda$ away from the Fermi surface.

Evaluating the Hall conductivity this way, there are two contributions to $\sigma_{xy}^{\rm H}$.
The first contribution comes from off-diagonal matrix elements of the renormalized velocity vertices. Subtracting the pure bare-velocity product already contained in Eq.~\eqref{eq:appIntrinsicVelocity} gives\footnote{Although the Kubo bubble may be drawn with one full vertex, the band-projected result is conveniently written in a symmetric form involving the corresponding left and right transport vertices.}
\begin{align}
 \label{eq:appVertexRaw}
    \sigma_{\beta\alpha}^{(v)}=&\, \frac{e^2\hbar}{2\pi}\int dE\,[-f'(E)]\sum_\lambda\sum_{\lambda'\neq\lambda}  \\ \notag
    & \times \int_{\bm k} \frac{(\Gamma_\beta)_{\lambda\lambda'}(\Gamma_\alpha)_{\lambda'\lambda}-(v_\beta)_{\lambda\lambda'}(v_\alpha)_{\lambda'\lambda}}{(E-\varepsilon_{\lambda'\bm k}+i\eta_{\lambda'\bm k})(E-\varepsilon_{\lambda\bm k}-i\eta_{\lambda\bm k})}.
\end{align}
The second correction is generated by the band-off-diagonal self-energy contained in the dressed propagators. To first order in this band-off-diagonal part of the self-energy one has
\begin{align}
    \bG^{R/A} =&\, \sum_\lambda\frac{P_{\lambda\bm k}}{E-\varepsilon_{\lambda\bm k}\pm i\eta_{\lambda\bm k}} 
    \notag \\ \notag &\, \mbox{}
    + \sum_{\lambda\neq\lambda'} \frac{P_{\lambda\bm k}\Sigma^{R/A}P_{\lambda'\bm k}}{(E-\varepsilon_{\lambda\bm k}\pm i\eta_{\lambda\bm k})(E-\varepsilon_{\lambda'\bm k}\pm i\eta_{\lambda'\bm k})}
    \\ &\, \mbox{} +\ldots.
 \label{eq:appGOffDiagonal}
\end{align}
Keeping one Fermi-surface band $\lambda$ and one band $\lambda'\neq\lambda$ away from the Fermi surface gives
\begin{align}
    \sigma_{\beta\alpha}^{(\Sigma)} =&\, \frac{e^2\hbar}{2\pi}\int dE\,[-f'(E)]\sum_\lambda\sum_{\lambda'\neq\lambda} \notag \\
    &\, \mbox{} \times \int_{\bm k}\frac{\mathcal N_{\beta\alpha}^{\lambda\lambda'}(\bm k,E)}{[(E-\varepsilon_{\lambda\bm k})^2+\eta_{\lambda\bm k}^2](E-\varepsilon_{\lambda'\bm k})}, \label{eq:appSelfEnergyRaw}
\end{align}
where
\begin{align}
    \mathcal N_{\beta\alpha}^{\lambda\lambda'}&=\left[(\Gamma_\beta)_{\lambda\lambda'}(\Sigma^R)_{\lambda'\lambda}+(\Sigma^A)_{\lambda\lambda'}(\Gamma_\beta)_{\lambda'\lambda}\right](\Gamma_\alpha)_{\lambda\lambda} \notag \\
    & +(\Gamma_\beta)_{\lambda\lambda}\left[(\Sigma^R)_{\lambda\lambda'}(\Gamma_\alpha)_{\lambda'\lambda}+(\Gamma_\alpha)_{\lambda\lambda'}(\Sigma^A)_{\lambda'\lambda}\right]. \label{eq:appSelfEnergyNumerator}
\end{align}
All self-energies and vertices in Eqs.~\eqref{eq:appSelfEnergyRaw} and~\eqref{eq:appSelfEnergyNumerator} are evaluated at the same energy $E$.

The principal-value part of Eq.~\eqref{eq:appVertexRaw} and the Hermitian real part of the band-off-diagonal self-energy are symmetric under $\alpha\leftrightarrow\beta$ at this order. The Hall response is therefore controlled by the imaginary part of the self energy $\eta(E)$, see Eq.\ (\ref{eq:appEtaDefinition}). Performing the energy integral in Eq.~\eqref{eq:appVertexRaw} gives
\begin{align}
 \label{eq:appVertexHall}
    \sigma_{\beta\alpha}^{(v),{\rm H}}=&\, e^2\hbar\sum_\lambda\sum_{\lambda'\neq\lambda}\int_{\bm k}[-f'(\varepsilon_{\lambda\bm k})] \\ \notag
    &\, \mbox{} \times\frac{\Im\!\left[(\Gamma_\beta)_{\lambda\lambda'}(\Gamma_\alpha)_{\lambda'\lambda}-(v_\beta)_{\lambda\lambda'}(v_\alpha)_{\lambda'\lambda}\right]}{\varepsilon_{\lambda'\bm k}-\varepsilon_{\lambda\bm k}}.
\end{align}
Similarly, Eq.~\eqref{eq:appSelfEnergyRaw} becomes
\begin{align}
 \label{eq:appSelfEnergyHall}
    \sigma_{\beta\alpha}^{(\Sigma),{\rm H}}=&\, -e^2\hbar\sum_\lambda\sum_{\lambda'\neq\lambda}\int_{\bm k}\frac{[-f'(\varepsilon_{\lambda\bm k})]}{\varepsilon_{\lambda'\bm k}-\varepsilon_{\lambda\bm k}} \\ \notag
    &\, \mbox{} \times\Im\!\left\{\frac{\eta_{\lambda\lambda'}}{\eta_{\lambda\bm k}}\left[(\Gamma_\alpha)_{\lambda'\lambda}(\Gamma_\beta)_{\lambda\lambda}-(\Gamma_\beta)_{\lambda'\lambda}(\Gamma_\alpha)_{\lambda\lambda}\right]\right\},
\end{align}
where
\begin{equation}
    \eta_{\lambda\lambda'}=\bra{u_{\lambda\bm k}}\eta(\varepsilon_{\lambda\bm k})\ket{u_{\lambda'\bm k}}.
 \label{eq:appEtaOffDiagonal}
\end{equation}
The complete leading non-crossing Hall response is therefore
\begin{equation}
   \sigma_{\beta\alpha}^{\rm H}=\sigma_{\beta\alpha}^{\rm B}+\sigma_{\beta\alpha}^{(v),{\rm H}}+\sigma_{\beta\alpha}^{(\Sigma),{\rm H}}+O\!\left(\frac{e^2}{\hbar}\frac{\hbar}{E_F\tau}\right).
 \label{eq:appHallBookkeeping}
\end{equation}

\subsection{Combined response and transport decomposition}

The two Fermi-surface corrections can be combined by introducing the orbital-diagonal matrix
\begin{align}
    \mathcal U_{\lambda,\alpha}&=(\Gamma_{\lambda\bm k,\alpha} - v_{\lambda\bm k,\alpha}) - \Gamma_{\lambda\bm k,\alpha} \frac{\eta(\varepsilon_{\lambda\bm k})}{\eta_{\lambda\bm k}},
 \label{eq:appCombinedU}
\end{align}
where
\begin{equation}
    v_{\lambda\bm k,\alpha}=(v_\alpha)_{\lambda\lambda},\qquad \Gamma_{\lambda\bm k,\alpha}=(\Gamma_\alpha)_{\lambda\lambda}.
 \label{eq:appDiagonalVertices}
\end{equation}
The first term in Eq.~\eqref{eq:appCombinedU} is the non-crossing ladder correction. The second represents a band-off-diagonal self-energy insertion multiplied by the diagonal transport vertex on the Fermi-surface band. Both remain finite as the disorder strength tends to zero because the Fermi-surface retarded--advanced pole supplies the compensating factor of $\tau$.
Adding Eqs.~\eqref{eq:appVertexHall} and~\eqref{eq:appSelfEnergyHall} gives
\begin{align}
    \sigma_{\beta\alpha}^{{\rm H}(v)}+\sigma_{\beta\alpha}^{{\rm H}(\Sigma)}&=e^2\hbar\sum_\lambda\sum_{\lambda'\neq\lambda}\int_{\bm k}[-f'(\varepsilon_{\lambda\bm k})] 
  \frac{\Im\,\mathcal Q_{\beta\alpha}^{\lambda\lambda'}(\bm k)}{\varepsilon_{\lambda'\bm k}-\varepsilon_{\lambda\bm k}},
 \label{eq:appCombinedFermiSurface}
\end{align}
with
\begin{align}
    \mathcal Q_{\beta\alpha}^{\lambda\lambda'}&=\left[v_\beta+\mathcal U_{\lambda,\beta}\right]_{\lambda\lambda'}\left[v_\alpha+\mathcal U_{\lambda,\alpha}\right]_{\lambda'\lambda} 
  -(v_\beta)_{\lambda\lambda'}(v_\alpha)_{\lambda'\lambda}.
 \label{eq:appCombinedQ}
\end{align}
Combining this result with Eq.~\eqref{eq:appSeaOffBerry} yields
\begin{align}
    \sigma_{\beta\alpha}^{\rm H}&=\sigma_{\beta\alpha}^{\rm B}+e^2\hbar\sum_\lambda\sum_{\lambda'\neq\lambda}\int_{\bm k}[-f'(\varepsilon_{\lambda\bm k})]\frac{\Im\,\mathcal Q_{\beta\alpha}^{\lambda\lambda'}(\bm k)}{\varepsilon_{\lambda'\bm k}-\varepsilon_{\lambda\bm k}}.
 \label{eq:appCompleteHall}
\end{align}
This is the complete leading weak-disorder Hall response for the model with uncorrelated on-site disorder.

Expanding Eq.~\eqref{eq:appCombinedQ} separates the Fermi-surface correction into the conventional anomalous-distribution, side-jump, and pair-skew pieces:
\begin{align}
    \sigma_{\beta\alpha}^{\rm a.d.}&=e^2\hbar\sum_\lambda\sum_{\lambda'\neq\lambda}\int_{\bm k}[-f'(\varepsilon_{\lambda\bm k})]\frac{\Im\!\left[(\mathcal U_{\lambda,\beta})_{\lambda\lambda'}(v_\alpha)_{\lambda'\lambda}\right]}{\varepsilon_{\lambda'\bm k}-\varepsilon_{\lambda\bm k}}, \label{eq:appAnomalousDistribution}\\
    \sigma_{\beta\alpha}^{\rm s.j.}&=e^2\hbar\sum_\lambda\sum_{\lambda'\neq\lambda}\int_{\bm k}[-f'(\varepsilon_{\lambda\bm k})]\frac{\Im\!\left[(v_\beta)_{\lambda\lambda'}(\mathcal U_{\lambda,\alpha})_{\lambda'\lambda}\right]}{\varepsilon_{\lambda'\bm k}-\varepsilon_{\lambda\bm k}}, \label{eq:appSideJump}\\
    \sigma_{\beta\alpha}^{\rm p.s.}&=e^2\hbar\sum_\lambda\sum_{\lambda'\neq\lambda}\int_{\bm k}[-f'(\varepsilon_{\lambda\bm k})]\frac{\Im\!\left[(\mathcal U_{\lambda,\beta})_{\lambda\lambda'}(\mathcal U_{\lambda,\alpha})_{\lambda'\lambda}\right]}{\varepsilon_{\lambda'\bm k}-\varepsilon_{\lambda\bm k}}. \label{eq:appPairSkew}
\end{align}
Here, the terms $\sigma_{\beta\alpha}^{\rm a.d.}$ and $\sigma_{\beta\alpha}^{\rm s.j.}$ are no longer antisymmetric in the indices. Instead, one has the relation
\begin{equation}
    \sigma_{\beta\alpha}^{\rm a.d.}=-\sigma_{\alpha\beta}^{\rm s.j.},
 \label{eq:appADSJRelation}
\end{equation}
ensuring antisymmetry of the full anomalous Hall conductivity $\sigma_{\alpha\beta}^{\rm H}$.

The pair-skew term in Eq.~\eqref{eq:appPairSkew} is a two-impurity contribution, which contributes to $O(\tau^0)$ in the weak-disorder limit. In general, there are additional pair-skew-scattering contributions, which also contribute to the Hall conductivity to order $O(\tau^0)$ in the weak-disorder limit. These contributions involve the crossed $X$ and $\Psi$ diagrams \cite{Ado2015}. For a tight-binding graph with uncorrelated on-site disorder, these additional pair-skew-scattering contributions are absent, as shown in detail in App.\ \ref{app:crossed}. In models with non-Gaussian disorder, there is an additional skew-scattering contribution \cite{SinitsynEtAl2007}, which involves scattering from a single impurity only and already contributes to $\sigma^{\rm H}_{xy}$ to order $\tau$. Such contribution is absent from the Gaussian model we consider here.

\section{Crossed diagrams for uncorrelated on-site disorder}
\label{app:crossed}

For a generic disorder ensemble, diagrams containing one pair of crossed impurity lines, as illustrated in the lower row of Fig.~\ref{fig:diagrams}, contribute to the Hall conductivity at the same order $O[(\hbar/E_F\tau)^0]$ as the non-crossing side-jump and pair-skew terms~\cite{Ado2015}. The diagrams in the bottom row of Fig.\ \ref{fig:diagrams} are referred to as the $X$ diagram and two $\Psi$ diagrams. In this appendix, we show that their contribution to the antisymmetric part of the conductivity tensor vanishes at this order for the model of uncorrelated on-site disorder considered here (see Eq.~\eqref{eq:diagdisorder}).

We use the shorthand notation $\bG_{\bm k}^{R/A}=\bG^{R/A}(\bm k,E)$. The right and left ladder-dressed vertices are $\Gamma_\alpha^R$ and $\Gamma_\alpha^L$, respectively, with
\begin{align}
    \Gamma_\alpha^R&=v_\alpha + \sum_q\gamma_qP_q\int_{\bm k}\bG^R\Gamma_\alpha^R\bG^AP_q, \label{eq:appRightVertex}\\
    \Gamma_\alpha^L&=v_\alpha + \sum_q\gamma_qP_q\int_{\bm k}\bG^A\Gamma_\alpha^L\bG^RP_q. \label{eq:appLeftVertex}
\end{align}
At the leading Drude order needed in the crossed diagrams, their band-diagonal projections are real and coincide. We denote the common transport vertex by
\begin{equation}
 \mathcal V_{\lambda\alpha}(\bm k,E)=\bra{u_{\lambda\bm k}}\Gamma_\alpha^{R/L}(\bm k,E)\ket{u_{\lambda\bm k}}.
 \label{eq:appTransportVertex}
\end{equation}

\subsection{The $X$ diagram}

The $X$-diagram contribution is
\begin{align}
    \sigma_{\beta\alpha}^{X}&=\frac{e^2\hbar}{2\pi}\int dE\,[-f'(E)]\sum_{q,q'}\gamma_q\gamma_{q'} \notag \\
    & \times \int_{\bm k, \bm k', \bm p} \bra{q'}\bG_{\bm k}^A\Gamma_\beta^L(\bm k)\bG_{\bm k}^R\ket q\bra{q'}\bG_{\bm k'}^R\Gamma_\alpha^R(\bm k')\bG_{\bm k'}^A \ket q  \notag \\
    & \quad  \times \bra q\bG_{\bm p}^R\ket{q'}\bra q\bG_{\bm k+\bm k'-\bm p}^A\ket{q'},
 \label{eq:appXfull}
\end{align}
where the states $\ket{q}$ are defined in the orbital basis.
The dominant weak-disorder contribution is obtained by projecting each retarded--advanced pair at a common momentum onto one band. Then
\begin{align}
    \bG_{\bm k}^A\Gamma_\alpha^{L/R}\bG_{\bm k}^R \longrightarrow\sum_\lambda\frac{P_{\lambda\bm k}\mathcal V_{\lambda\alpha}}{(E-\varepsilon_{\lambda\bm k})^2+\eta_{\lambda\bm k}^2}.
 \label{eq:appCrossedProjection}
\end{align}
Each Fermi-surface retarded--advanced pair supplies one factor of $\tau$, while the two impurity contractions supply $\tau^{-2}$; the projected diagram is therefore of order $\tau^0$. Using Eq.~\eqref{eq:appCrossedProjection},
\begin{align}
    \sigma_{\beta\alpha}^{X,(0)} =&\, \frac{e^2\hbar}{2\pi}\int dE\,[-f'(E)]\sum_{q,q'}\gamma_q\gamma_{q'}\sum_{\lambda,\lambda'} \notag \\
    &\, \mbox{} \times \int_{\bm k, \bm k', \bm p}  \frac{\braket{q'}{u_{\lambda\bm k}}\mathcal V_{\lambda\beta}(\bm k,E)\braket{u_{\lambda\bm k}}{q}}{(E-\varepsilon_{\lambda\bm k})^2+\eta_{\lambda\bm k}^2} 
    \notag \\ &\, \mbox{} \times
    \frac{\braket{q'}{u_{\lambda'\bm k'}}\mathcal V_{\lambda'\alpha}(\bm k',E)\braket{u_{\lambda'\bm k'}}{q}}{(E-\varepsilon_{\lambda'\bm k'})^2+\eta_{\lambda'\bm k'}^2} \notag\\
    &\, \mbox{} \times\bra q\bG_{\bm p}^R\ket{q'}\bra q\bG_{\bm k+\bm k'-\bm p}^A\ket{q'}.
 \label{eq:appXprojected}
\end{align}
Equation \eqref{eq:appXprojected} is invariant under $\alpha\leftrightarrow\beta$, $\bm k\leftrightarrow\bm k'$, and $\lambda\leftrightarrow\lambda'$. Consequently, $\sigma_{\beta\alpha}^{X,(0)}=\sigma_{\alpha\beta}^{X,(0)}$, so that the antisymmetric part
\begin{equation}
  \sigma_{\beta\alpha}^{X,(0),\rm H}=0
 \label{eq:Xzero}
\end{equation}
at $O[(\hbar/E_F\tau)^0]$.

\subsection{The $\Psi$ diagrams}

The two $\Psi$ diagrams differ by whether the propagators connecting the crossed impurity lines are both retarded or both advanced. Their sum is
\begin{align}
    \sigma_{\beta\alpha}^{\Psi} =&\, \frac{e^2\hbar}{2\pi}\int dE\,[-f'(E)]\sum_{q,q'}\gamma_q\gamma_{q'} \int_{\bm k, \bm k', \bm p} \notag \\
    &\, \mbox{} \times \bra{q'}\bG_{\bm k}^A\Gamma_\beta^L(\bm k)\bG_{\bm k}^R\ket q \bra q\bG_{\bm k'}^R\Gamma_\alpha^R(\bm k')\bG_{\bm k'}^A\ket{q'} \notag\\
    &\, \mbox{} \times\left[\bra q\bG_{\bm p}^R\ket{q'}\bra{q'}\bG_{\bm p - \bm k + \bm k'}^R\ket q 
    \right. \nonumber \\ &\, \ \ \ \ \mbox{} \left.
    +\bra{q'}\bG_{\bm p}^A\ket q\bra q\bG_{\bm p - \bm k + \bm k'}^A\ket{q'}\right],
 \label{eq:appPsifull}
\end{align}
After projecting the two retarded--advanced pairs onto the Fermi surface,
\begin{align}
    \sigma_{\beta\alpha}^{\Psi,(0)} =&\, \frac{e^2\hbar}{2\pi}\int dE\,[-f'(E)]\sum_{q,q'}\gamma_q\gamma_{q'}\sum_{\lambda,\lambda'} \notag \\
    &\, \mbox{} \times \int_{\bm k, \bm k', \bm p} 
    \frac{\braket{q'}{u_{\lambda\bm k}}\mathcal V_{\lambda\beta}(\bm k,E)\braket{u_{\lambda\bm k}}{q}}{(E-\varepsilon_{\lambda\bm k})^2+\eta_{\lambda\bm k}^2}
    \nonumber \\ &\, \mbox{} \times
    \frac{\braket q{u_{\lambda'\bm k'}}\mathcal V_{\lambda'\alpha}(\bm k',E)\braket{u_{\lambda'\bm k'}}{q'}}{(E-\varepsilon_{\lambda'\bm k'})^2+\eta_{\lambda'\bm k'}^2} \notag\\
    &\, \mbox{} \times\left[\bra q\bG_{\bm p}^R\ket{q'}\bra{q'}\bG_{\bm p - \bm k + \bm k'}^R\ket q 
    \right. \nonumber \\ &\, \mbox{} \ \ \ \ \left. 
    +\bra{q'}\bG_{\bm p}^A\ket q\bra q\bG_{\bm p - \bm k + \bm k'}^A\ket{q'}\right].
 \label{eq:appPsiprojected}
\end{align}
The transformation $\alpha\leftrightarrow\beta$, $\bm k\leftrightarrow\bm k'$, $\lambda\leftrightarrow\lambda'$, and $q\leftrightarrow q'$ leaves Eq.~\eqref{eq:appPsiprojected} invariant, so that we again conclude that the contribution to the antisymmetric component of the conductivity tensor
\begin{equation}
    \sigma_{\beta\alpha}^{\Psi,(0),\rm H}=0
 \label{eq:Psizero}
\end{equation}
at $O[(\hbar/E_F\tau)^0]$.

We note that the conclusion that $\sigma_{\beta\alpha}^{X,(0),\rm H} = \sigma_{\beta\alpha}^{\Psi,(0),\rm H} = 0$ relies on the independent orbital projectors in  Eq.~\eqref{eq:diagdisorder}, which appears because of the absence of correlations between the disorder potentials at different lattice sites. It does not apply to a random potential that has correlations between different orbitals, as shown in the examples considered in Ref.\ \cite{Ado2015}.

\section{Symmetry action on the embedding-sensitivity vectors}
\label{app:symmetry}

Let $g$ be a spatial symmetry with momentum-space action $\bm k\mapsto R_g\bm k$, orbital permutation $q\mapsto gq$, and the induced band permutation $\lambda\mapsto g\lambda$. The energies and orbital weights obey
\begin{align}
    \varepsilon_{g\lambda}(R_g\bm k)&=\varepsilon_\lambda(\bm k), \label{eq:energysym}\\
    w_{g\lambda,gq}(R_g\bm k)&=w_{\lambda q}(\bm k). \label{eq:weightsym}
\end{align}
Differentiating the first relation gives
\begin{equation}
    \nabla_{R_g\bm k}\varepsilon_{g\lambda}(R_g\bm k)=R_g\nabla_{\bm k}\varepsilon_\lambda(\bm k). \label{eq:velsym}
\end{equation}
In two dimensions, the matrix $J$ representing $\hat{\bm z}\times$ satisfies $JR_g=\det(R_g)R_gJ$. Changing variables in Eq.~\eqref{eq:xi}, summing over all bands at energy $E$, and relabeling $g\lambda\to\lambda$ gives Eq.~\eqref{eq:xisymmetry}.

If a subgroup $G_q$ fixes the orbital label $q$, then $\bm\xi_q$ lies in the common invariant subspace of $D_g=\det(R_g)R_g$ for all $g\in G_q$. The vector must vanish when this common subspace is trivial. If a symmetry permutes an orbital orbit $\mathcal O=\{q,gq,\ldots\}$, the collection $\{\bm\xi_q:q\in\mathcal O\}$ transforms covariantly. The orbit sum may vanish even when the individual vectors do not.

\begin{figure}[t!]
    \centering    \includegraphics[width=0.95\columnwidth]{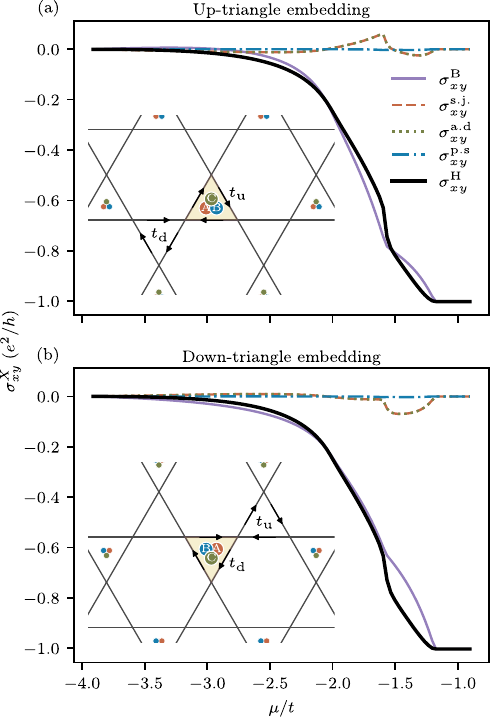}
    \caption{Embedding dependence of the anomalous Hall conductivity of a flux-threaded kagome model, see Eq.~\eqref{eq:kagome-hamiltonian}. The panels show the
    Hall-conductivity contributions for the periodic gauge, using unit cells centered on an
    (a) up-pointing and (b) down-pointing triangle. The insets indicate the
    corresponding positions of the three orbitals within each unit cell.  The figure shows the embedding-dependent individual contributions to the Hall conductivity as well as the embedding-independent total anomalous Hall response $\sigma^{\text{H}}_{xy}$.}
    \label{fig:kagome-embedding}
\end{figure}

\section{Numerical procedure}
\label{app:numerics}

We evaluate the non-crossing Hall response following the procedure described in Sec.~\ref{sec:disorder}. For every embedding and energy, we solve Eqs.~\eqref{eq:scba} and~\eqref{eq:BSmain}, insert the resulting self-energy and dressed vertices into Eqs.~\eqref{eq:KSsea} and~\eqref{eq:fullRAbubble}, and apply the intra- and inter-band velocity decomposition introduced in the main text.

Energies are measured in units of the characteristic hopping amplitude $t$, the lattice constant is set to $1$, and momenta are expressed in the corresponding dimensionless units. We consider the case of the on-site disorder that has the same variance on all orbitals,
\begin{equation}
    \gamma_q = \gamma,
    \qquad \forall q,
    \label{eq:gamma_equal}
\end{equation}
where $\gamma$ has units of energy squared.

The disorder-induced broadening is parametrically of order $\gamma/t$. Accordingly, we choose the energy spacing for the Fermi-sea integration as
\begin{equation}
    \Delta E = \alpha \frac{\gamma}{t},
    \label{eq:energy_grid}
\end{equation}
where $\alpha$ is a dimensionless numerical parameter. The Brillouin zone is sampled on a uniform $N_k\times N_k$ mesh, with the momentum
spacing chosen such that
\begin{equation}
    \Delta k \simeq \frac{\Delta E}{t}
    = \alpha\frac{\gamma}{t^2}.
    \label{eq:momentum_grid}
\end{equation}
Here $\Delta k$ is dimensionless because the lattice constant has been set to unity. 

All results are obtained at zero temperature, $k_{\rm B}T=0$. The self-consistent Born equation for the self-energy is iterated to a relative tolerance of $10^{-8}$. In our numerical calculations, we use $\alpha=0.1$ and $\gamma/t^2=0.05$. Convergence is checked and estimated by reducing $\alpha$ and $\gamma/t^2$. At fixed $\alpha=0.2$, changing $\gamma/t^2$ from $0.05$ to $0.025$ changes $\sigma_{xy}^{\rm H}$ by at most $1.41\times10^{-2}\,e^2/h$, corresponding to a maximum relative difference of $1.41\%$. At fixed $\gamma/t^2=0.05$, changing $\alpha$ from $0.2$ to $0.1$ changes $\sigma_{xy}^{\rm H}$ by at most $6.42\times10^{-5}\,e^2/h$, corresponding to a maximum relative difference of $0.0064\%$. Adding these two estimates conservatively, we assign a maximum relative uncertainty of approximately $1.5\%$ to the reported conductivity.

\subsection{Flux-threaded kagome model}

As an additional numerical example, we consider a kagome lattice with three sites, $A$, $B$, and $C$, per unit cell. Complex nearest-neighbor hopping amplitudes generate finite fluxes through the elementary up and down triangles. We define the three bond momenta
\begin{equation}
 k_1=k_x,\quad
 k_2=-\frac{k_x}{2}-\frac{\sqrt{3}k_y}{2},\quad
 k_3=-\frac{k_x}{2}+\frac{\sqrt{3}k_y}{2},
\end{equation}
and the hopping amplitudes
\begin{equation}
 t_{\mathrm u}
 =
 r_{\mathrm u}
 \exp\left[i\frac{\pi+\phi_{\mathrm u}}{3}\right], \quad
 t_{\mathrm d}
 =
 r_{\mathrm d}
 \exp\left[i\frac{\pi+\phi_{\mathrm d}}{3}\right],
\end{equation}
where the subscripts ``u" and ``d" refer to the up and down triangles, respectively.

In Fig.~\ref{fig:kagome-embedding} we compare the Hall conductivity calculated using the two embeddings of the same kagome graph. In the two choices, the three sites belonging to each upward- or downward-pointing triangle are assigned to the center of that triangle. These two choices correspond to the two different conventions for the unit cell and represent the two natural periodic gauges.
In the first choice, corresponding to the up-triangle embedding shown in Fig.~\ref{fig:kagome-embedding}(a), the Bloch Hamiltonian is
\begin{equation}
 h_0(\bm{k})
 = \begin{pmatrix}
 0
 &
 t_{\mathrm u}+t_{\mathrm d}e^{-ik_1}
 &
 t_{\mathrm u}^{*}+t_{\mathrm d}^{*}e^{ik_2}
 \\
 t_{\mathrm u}^{*}+t_{\mathrm d}^{*}e^{ik_1}
 &
 0
 &
 t_{\mathrm u}+t_{\mathrm d}e^{-ik_3}
 \\
 t_{\mathrm u}+t_{\mathrm d}e^{-ik_2}
 &
 t_{\mathrm u}^{*}+t_{\mathrm d}^{*}e^{ik_3}
 &
 0
 \end{pmatrix}.
 \label{eq:kagome-hamiltonian}
\end{equation}
The down-triangle embedding in Fig.~\ref{fig:kagome-embedding}(b) is obtained from Eq.~\eqref{eq:kagome-hamiltonian} by the unitary transformation~\eqref{eq:hchange} with
\begin{equation}
U_{\bm{k}} = e^{i(k_1+2k_2)/3}
    \begin{pmatrix}
        e^{-ik_2} & 0 & 0 \\
        0 & e^{ik_3} & 0 \\
        0 & 0 & 1
    \end{pmatrix}
\end{equation}
in the $(A,B,C)$ basis. Since $k_1+k_2+k_3=0$, this transformation makes the down-triangle hoppings momentum independent, while the up-triangle hoppings acquire the corresponding Bloch phases.

For the numerical results presented in Fig.~\ref{fig:kagome-embedding}, we take
\begin{equation}
 r_{\mathrm u}=r_{\mathrm d}=t,\qquad
 \phi_{\mathrm u}=1,\qquad
 \phi_{\mathrm d}=\frac{1}{2}.
\end{equation}
For this choice of parameters, the kagome model is a gapped Chern insulator. For a chemical potential $\mu$ in the spectral gap, the Hall conductivity is quantized and given by the band contribution $\sigma_{xy}^{\rm B}$, independent of the embedding. The Hall conductivity remains embedding-independent if $\mu$ is in the valence band, but the individual contributions to the Hall conductivity acquire a dependence on the embedding choice, consistent with the conclusions of the main text.

\bibliography{bib-AHE}

\end{document}